\documentclass[
preprint,epsfig,showkeys,amsmath,amssymb,aps,physrev,floatfix,]{revtex4}
\usepackage{graphicx}
\usepackage{bm}

\begin{document}

\title{Poincar\'e asymptotic expansion in black hole theory}

\author{Giampiero Esposito}
\affiliation{Dipartimento di Fisica 
``Ettore Pancini'', Complesso Universitario di Monte S. Angelo,
Via Cintia Edificio 6, 80126 Napoli, Italy \\
\quad Istituto Nazionale di Fisica Nucleare, Sezione di Napoli, 
Complesso Universitario di Monte S. Angelo,
Via~Cintia Edificio 6, 80126 Napoli, Italy}

\affiliation{giampiero.esposito@na.infn.it\\ORCID: 0000-0001-5930-8366}
\author{Marco Refuto}%
\affiliation{Dipartimento di Fisica 
``Ettore Pancini'', Complesso Universitario di Monte S. Angelo,
Via Cintia Edificio 6, 80126 Napoli, Italy \\
\quad Istituto Nazionale di Fisica Nucleare, Sezione di Napoli, 
Complesso Universitario di Monte S. Angelo,
Via~Cintia Edificio 6, 80126 Napoli, Italy}
 \affiliation{marco.refuto@unina.it \\ ORCID: 0009-0007-6563-7909}

\begin{abstract}
In studying the dynamics of fields in black hole theory, the method of separation of 
variables makes it possible to isolate the radial part of the full solution 
in many important physical cases. This occurs by virtue of the existence of 
the principal tensor in Petrov-D metrics. We first review this mathematical result in 
order to introduce several cases where it is possible to study the radial solution 
via the Poincaré asymptotic series expansion, a tool exploited in 
recent work by the authors in order to investigate the 
behaviour of the field at spacelike infinity, a 
point in the neighbourhood of which 
only approximate solutions are computable by virtue of its irregular 
nature. We obtain a series which can be computed to any degree of accuracy, 
allowing for a deeper analysis of this challenging space-time region. 
An application to quasinormal modes is eventually provided.
\end{abstract}

                              
\maketitle

\tableofcontents

\section{Introduction}

Black hole physics \cite{HE1973, Chan1984} is a never-ending reservoir of tests 
for our knowledge of physics. Even if such a topic covers several aspects of 
our modern theories, it is possible to recognize two main features at its basis. 
The first goes under the name of \textit{cosmic censor conjecture} (see Refs 
\cite{WALD1974, JW1977} for informal discussions): the complete gravitational 
collapse of a body always results in a black hole rather than a naked (visible 
for a distant observer) singu\-la\-ri\-ty. The other one is the \textit{no hair} 
theorem: all black hole solutions to the Einstein-Maxwell equations are uniquely 
characterized by mass, electric charge and angular momentum. Even if more and 
more studies are questioning the validity of such a theorem in various scenarios, 
it has a well-established foundation and, at least to a first approximation, we 
can say that these three parameters identify different kinds of black holes, 
depending on which of them is set equal to zero (we are not considering, e.g., 
the soft hair on black holes \cite{HPS2016} which arises from the BMS group 
in asymptotically flat space-times). 
Space-time properties have been studied extensively for decades. The same happens 
for the dynamics of fields on them, even if only recently several results are fully 
understood thanks to mathematical and computational progress.  For instance, in his pioneer 
analysis of the radial wave function for a massless scalar field in Schwarzschild space-time, 
Persides \cite{Persides1976a,Persides1976b} was able to find only local solutions 
near the singular points of the equation through an approximation scheme. Today, 
instead, it is understood that the Heun theory of differential equations covers 
this and many other cases in a complete mathematical framework.
One of the most interesting properties of the field equations of motion in black-hole 
space-times is that they are separable: there exists a coordinate system in which the 
solution can be expressed as a product of four functions of a single variable. While it 
could be not surprising in the Schwarzschild case (its high degree of symmetry makes 
the integration of the equations straightforward), it holds also for the Kerr black hole 
and its several generalizations. This is nothing but a consequence of the concept of 
hidden symmetries (cf. Ref. \cite{Frolov2017kze} for a detailed discussion, 
we will cover some aspects in following sections): symmetries which do not belong to 
the space-time, but emerge only when one looks at the dynamics. Black hole metrics, 
which are all Petrov-D type, admit the existence of a peculiar tensor called the 
principal tensor. Such an object allows for the separability of the Klein-Gordon and 
Dirac equations and their complete integrability. In this paper, we rely on this powerful 
result in order to apply to many other cases the Poincar\'e asymptotic series expansion: 
a technique we exploited in our previous paper \cite{ER} in order to study the behaviour 
at spacelike infinity for a massive scalar field in Schwarzschild space-time. We were 
able to compute with arbitrary precision the solution instead of relying on approximation 
schemes considered in the literature. We computed also the complete spectrum of the field 
at spacelike infinity, integrating the single Fourier's mode over all frequencies. 
In conclusion, we solved the Zerilli equation with our method, finding full agreement 
with the results found in the literature, for instance in Ref. \cite{Hamaide2023}. 
Such a success in the easiest black hole description provided by Schwarzschild space-time 
suggests going beyond this model. Testing the method in several different scenarios 
and connecting it with already known results (usually obtained via Green functions or 
partial wave decomposition), could be useful in order to add a new tool to the 
investigations of fields' behaviour in curved space-times. 
\newline
In Sec. \ref{Poincare asymptotic series expansion} we review the Poincar\'e theory we 
already covered in our previous paper \cite{ER}. It aims at resolving an ordinary 
differential equation with variable coefficients, giving its asymptotic expansion.
\newline
In Sec. \ref{Stationary, axial-symmetric black holes} we focus the attention on the 
space-times where we study the dynamics of fields. First, we give a rather lengthy 
description of properties of the Kerr-Newman space-time, since it provides an 
interesting scenario more general than Schwarzschild but at the same time easily 
comparable to it. We then focus on the concept of {\it hidden symmetries} and the 
{\it principal tensor}: it makes the Klein-Gordon and Dirac equations solvable 
via separation of variables for black hole metrics, allowing us to exploit the method 
shown in the previous section. We consider the Kerr-NUT-A(d)S space-time, where the 
NUT parameter and the cosmological constant are present. In conclusion, the most 
general four-dimensional electrovacuum solution of Einstein's equations which is 
stationary, axisymmetric and Petrov-D, i.e., the Pleba\'nski-Demia\'nski metric, is shown. 
\newline
In Sec. \ref{A massive scalar field in Kerr-Newman space-time} and Sec. \ref{A massive 
fermionic field in Kerr-Newman space-time} we study, for a massive field in 
Kerr-Newman space-time, the Klein-Gordon and Dirac equation, respectively. After a brief
introduction and comparison with what is already known in the literature we compute 
the radial solution at spacelike infinity through the Poincar\'e series, considering 
also the full spectrum analysis for a Klein-Gordon field.
\newline
In Sec. \ref{Spin-weighted solutions of Teukolsky equation}, we cover the case 
of a massless spin-weighted field in Kerr space-time, due to its wide importance in 
black hole physics. The central topic is therefore the Teukolsky equation.
In Sec. \ref{An application to quasinormal modes}, we provide a preliminary 
investigation of the Poincar\'e series expansion for the computation of quasinormal 
modes for a massless scalar field in Schwarzschild space-time. Among the several 
available techniques, we choose to exploit the Leaver's continuos functions method, 
showing that it is possible to compute quasinormal frequencies in this way, and
displaying the link with the reduction of radial wave equations to canonical form. 

\section{Poincar\'e's asymptotic expansion}
\label{Poincare asymptotic series expansion}

Instead of introducing directly the method, we would like to focus on the reasons 
which led us to consider such a framework as a tool of investigation of spacelike 
infinity, starting from the common choice adopted in the literature and then 
reviewing some topics we already discussed in Ref. \cite{ER}.

\subsection{Confluent Heun equations}

The mathematical description of the dynamics of fields in black hole physics relies 
often upon confluent Heun's equations \cite{Ronveaux1995} (or equations easily related 
to them), i.e., second-order ordinary differential equations of the form
\begin{equation}
\left[\frac{d^{2}}{dz^{2}}
+\left(\alpha+\frac{(\beta+1)}{z}
+\frac{(\gamma+1)}{(z-1)}\right)\frac{d}{dz} 
+ \left(\frac{\mu}{z}
+\frac{\nu}{(z-1)}\right) \right]Y(z)=0,
\label{confluent Heun equation in its general form}
\end{equation}
where the parameters $\mu$ and $\nu$ obey the relations
\begin{equation}
\mu=\frac{1}{2}(\alpha-\beta-\gamma-2 \eta+(\alpha - \gamma)\beta),
\end{equation}
\begin{equation}
\nu=\frac{1}{2}(\alpha+\beta+\gamma+2 \delta + 2 \eta
+(\alpha + \beta) \gamma),
\end{equation}
while $\alpha,\beta,\gamma,\delta,\eta$ are free (in general complex) parameters. 
We introduce a key quantity in the study of Heun equations, the $s$-rank: a 
classification of the irregularity of a singular point. Given a second-order linear 
ordinary differential equation, $y^{''}+P(z)y'+Q(z)y=0$, the $s$-rank of the singular 
point $z_0$ is defined as: $s=\text{max}(p-1,q/2)$, where $p,q$ are the pole order 
of $P(z)$ and $Q(z)$ at $z_0$, respectively. If the coefficients are analytic there, 
their $s$-rank is zero. The Eq. \eqref{confluent Heun equation in its general form} 
has two regular singular points at $z=0$, $z=1$ of $s$-rank $1$ and an 
irregular singular point at $z=\infty$ of
$s$-rank $2$, which in our cases corresponds to spacelike infinity. In the 
literature, only power-series solutions in the neighbourhood of singular points are 
known (it is however possible to extend them analytically under quite general assumptions), 
resulting in three sets of two linearly independent solutions. The~ones in the 
neighbourhood of $z=0$ and $z=1$ are called Frobenius solutions, while the ones in 
the neighbourhood of the point at infinity are known as
Thom\'e solutions. However, the last one is quite different from the former and deserves 
a different computation scheme. It is well known that the solution at infinity of 
Eq. \eqref{confluent Heun equation in its general form} is given by
\begin{equation}
Y_{\infty}(z)=e^{T_{n}(z)} \; z^{-\theta} \;
\sum_{k=0}^{\infty}\rho_{k}z^{-k},
\end{equation}
where $T_{n}$ is a polynomial function of $z$. Note that the sign of $\theta$ may 
change depending on the chosen parametrization. Upon~setting $T_{n}(z) \sim 1+z$ 
one obtains the so-called Thom\'e~solutions
\begin{equation}
Y_{\infty}(z)= C_{1}z^{-\theta_{1}} \sum_{k=0}^{\infty} \rho_{k}z^{-k}
+ C_{2}e^{\theta_{2}z} \; z^{-\theta_{3}} \;
\sum_{k=0}^{\infty}\sigma_{k}z^{-k}.
\end{equation}
By inserting them into Eq. \eqref{confluent Heun equation in its general form} it is 
possible to obtain a recurrence relation for the coefficients $\rho_{k},\sigma_{k}$. 
Note that $\theta_{1},\theta_{2},\theta_{3}$ are in principle different and complex. 
However, it is possible to relate them to  $\alpha,\beta,\gamma,\delta,\eta$ even though 
this strongly depends on the chosen~parametrization. The greatest difficulty lies 
precisely in the computation of $\rho_k,\sigma_k$, since they are solutions to 
second-order recurrence relations which often are impossible to solve exactly for 
every $k$, forbidding the possibility to compute every term of the series, even the 
first ones. For this reason, it is usually considered sufficient to work with the leading order.

\subsection{The Poincar\'e work on ordinary differential equations}

The Poincar\'e method \cite{Poincare1,Poincare2} has its root in this 
scenario, in order to offer an alternative route.
In particular, in Ref.~\cite{Poincare2}, Poincar\'e studied linear 
differential equations of the form
\begin{equation}
\sum_{k=0}^{n}P_{k}(x)\frac{d^{k}y}{dx^{k}}=0,
\label{(E1)}
\end{equation}
where the $P_{k}(x)$ are polynomials in the variable $x$. He knew from the work of Fuchs 
and Thom\'e that, apart from some exceptional cases, there exist 
$n$ functions of the following form:
\begin{equation}
e^{Q(x)} \; x^{a} \; \left(A_{0}+\frac{A_{1}}{x}
+\frac{A_{2}}{x^{2}}+...\right),
\label{(E2)}
\end{equation}
which satisfy formally Equation \eqref{(E1)}, where $Q(x)$ is a polynomial in $x$. 
The~series in Equation \eqref{(E2)} was then said to be a normal series. In~particular, 
if~the normal series is convergent, one says that Equation 
\eqref{(E1)} admits a normal~integral. 
Poincar\'e says that a divergent series 
$$
\sum_{k=0}^{\infty}\frac{A_{k}}{x^{k}},
$$
where the sum of the first $(n+1)$ terms is denoted by $S_{n}$, represents asymptotically 
a function $J: x \rightarrow J(x)$ if the expression 
$$
x^{n}(J(x)-S_{n}(x))
$$
approaches $0$ when $x$ tends to infinity:
\begin{equation}
\lim_{x \to \infty}x^{n}(J(x)-S_{n}(x))=0.
\label{(E3)}
\end{equation}
This means that, for~$x$ sufficiently large, one has the majorization
\begin{equation}
x^{n}(J(x)-S_{n}(x)) < \varepsilon.
\label{(E4)}
\end{equation}
The error
\begin{equation}
J(x)-S_{n}(x)=\frac{\varepsilon}{x^{n}}
\label{(E5)}
\end{equation}
made on taking (just) the first $(n+1)$ terms is then extremely 
small, and~also much smaller 
than the error made on taking just $n$ terms, which equals
\begin{equation}
J(x)-S_{n-1}(x)=\frac{A_{n}+\varepsilon}{x^{n}},
\label{(E6)}
\end{equation}
$\varepsilon$ being very small and $A_{n}$ being~finite. In theoretical physics
and applied mathematics, the modern literature still relies upon the Poincar\'e 
definition of asymptotic expansion. A sharply different definition, better suited
for studying the foundations of analysis, can be found instead in Ref. \cite{Dieu1980}.
A synthesis of such an alternative framework can be found in 
Appendix D of Ref. \cite{DEV2014}.

\subsection{The mathematical algorithm}

In order to apply these ideas to our cases, let us just consider a second-order ordinary 
differential equation with variable coefficients,
\begin{equation}
\left[\frac{d^{2}}{dr^{2}}+p(r)\frac{d}{dr}
+q(r)\right]R(r)=0.
\label{second order radial equation in the right form for poincare}
\end{equation}
It is possible to reduce it to a form where the first derivative is absent, usually known 
as Laguerre-Forsyth \cite{Wilc} canonical form \cite{Esposito2017}. Upon assuming 
\begin{equation}
    R(r)=\alpha(r)\beta(r),
\end{equation}
the method yields
\begin{equation}
    \alpha(r)={\rm exp} \left(-\frac{1}{2}\int p(r)dr \right),
\end{equation}
jointly with the second-order differential equation
\begin{equation}
\left[\frac{d^{2}}{dr^{2}}+J(r)\right]\beta(r)=0,
\label{(6.1)}
\end{equation}
where the potential term is given by
\begin{equation}
    J(r)=-\frac{1}{2}\frac{dp}{dr}
-\frac{1}{4}p^{2}(r)+q(r).
\end{equation}
The advantage of this form is that now we can consider the asymptotic expansion 
of the canonical potential $J(r)$, 
\begin{equation}
    \left(\frac{d^{2}}{dr^{2}}+\sum_{k=0}^{\infty}
\frac{A_{k}(\dots)}{r^{k}}\right)\beta(r)=0,
\label{(6.2)}
\end{equation}
where the $A_k$ coefficients are constant with respect to $r$ but in general might depend 
on several parameters denoted by dots, such as the charge or mass or the field, as well 
as the mass or angular momentum of the black hole. At~this stage, Eq. \eqref{(6.1)} 
suggests looking for $\beta_(r)$ at large $r$ in the form (setting 
$\varepsilon= \pm 1$ and considering 
$\zeta$ as a parameter to be determined as shown~below)
\begin{equation}
\beta(r) \sim e^{\varepsilon r \sqrt{-A_{0}}} \; r^\zeta
\; \left(1+\sum_{s=1}^{\infty}
\frac{B_{s}(\dots)}{r^{s}}\right).
\label{(6.2)}
\end{equation}
The insertion of Eq. \eqref{(6.2)} into Eq. \eqref{(6.1)} yields
\begin{equation}
\gamma_{0}+\frac{\gamma_{1}}{r}
+\frac{\gamma_{2}}{r^{2}}+\sum_{k=3}^{\infty}
\frac{\gamma_{k}}{r^{k}}=0.
\label{(6.3)}
\end{equation}
For this equation to be identically satisfied as $r \rightarrow \infty$, all 
$\gamma_{k}$ coefficients should vanish. Indeed, we find
\begin{equation}
\gamma_{0}=A_{0}-A_{0}=0,
\label{(6.4)}
\end{equation}
while
\begin{eqnarray}
  \gamma_{1}=&&2\zeta \varepsilon \sqrt{-A_0}+A_1,\\
    \gamma_{2}=&&(\zeta -1)(2\varepsilon B_1\sqrt{-A_0}+\zeta)+A_1B_1+A_2,
\end{eqnarray}
jointly with infinitely many other equations for all subsequent $\gamma_{k}$ coefficients. 
Since they should all vanish, we obtain linear algebraic equations for $B_{1},B_{2},...,B_{\infty}$. 
In~addition, since we look for a bounded solution at large $r$, we restrict ourselves 
to the $\varepsilon=-1$ case (the $\varepsilon=1$ mode cannot be avoided in studying, e.g., 
the behaviour near the horizon). We have therefore, for instance,
\begin{align}
    \zeta=&\frac{A_1}{2\sqrt{-A_0}}, 
    \label{zeta general expression}\\
    B_1=&\frac{\zeta(\zeta-1)+A_2}{2(\zeta-1)\sqrt{-A_0}-A_1}, 
    \label{B1 general expression}\\
    B_2=&\frac{2A_3\sqrt{-A_0}(\zeta-1)+\zeta^4-4\zeta^3+(2A_2+5)\zeta^2-2(2A_2+1)\zeta
-A_1A_3+A_2^2+2A_2}{(2\sqrt{-A_0}\zeta-2\sqrt{-A_0}-A_1)
(2\sqrt{-A_0}\zeta-4\sqrt{-A_0}-A_1)}.
\end{align}

We do not write explicitly further coefficients because they have a cumbersome form, however, 
they can be easily computed with symbolic algebraic programs, such as Mathematica or Maple. 
Note that the $B_s$ coefficients depend on the same parameters of $A_k$, obviously in a 
different manner. We report for completeness the full radial solution, given by 
\begin{equation}
    R(r)=\mathcal{N}\alpha(r)e^{- r \sqrt{-A_{0}}} \; r^\zeta
\; \left(1+\sum_{s=1}^{\infty}
\frac{B_{s}(\dots)}{r^{s}}\right),
\label{full radial asymptotic solution}
\end{equation}
where $\mathcal{N}$ is a normalization constant. In light of the above, the algorithm 
provided is well-suited for computer programs and makes it possible to obtain a solution with the 
desired level of accuracy, going beyond the leading order (which is usually similar to 
the case $B_s=0 $ for every $s$) so often considered in the literature.

\section{Stationary and axisymmetric black holes}
\label{Stationary, axial-symmetric black holes}

In this section we first review several properties of the Kerr-Newman space-time, i.e. a 
stationary and axisymmetric space-time offering a description of a charged and rotating 
black hole. The black hole rotation breaks up the spherical symmetry, while the presence 
of a charge modifies the scenario in a very predictable way. We aim at offering a 
physically-grounded motivation on why we cannot freely vary the parameters in the final 
solution: the constraints are intimately rooted in the foundations of general relativity. 
After such a digression, we enlarge the scope of our investigations briefly covering the 
Kerr-NUT-A(d)S space-time and its charged (and accelerating) solution, given by the 
Pleba\'nski-Demia\'nski space-time. We stress that the reliability of out method 
relies on the possibility to separate variables in the equation of motion and, in all  
these space-times, it is possible thanks to the existence of an object called principal tensor, 
making room for the so-called hidden symmetries. We just mention the basic construction, 
following mainly Ref. \cite{Frolov2017kze}, in order to have a solid ground for the next section.

\subsection{The Kerr-Newman black hole}
\label{The ker newman black hole, subsection}

The simplest mathematical description of a black hole is given by the Schwarzschild 
space-time \cite{SCH1916}: a static, spherically symmetric vacuum solution where 
charge and angular momentum are absent. Its generalization to the presence of an 
electromagnetic field was given independently by Reissner \cite{REISSNER1916} and 
Nordstrom \cite{NORDSTROM1918}. Another family of solutions was first discovered by Kerr 
\cite{Kerr1963}, imposing the condition of algebraic speciality on top of stationarity 
and axial symmetry of the Einstein equations, obtaining the description of a rotating black 
hole. Its charged counterpart was obtained by Newman \textit{et al.} \cite{Newman1965} 
shortly after. In agreement with the no hair theorem, the Kerr-Newman metric ascribes to a 
black hole all of the three parameters which are necessary and sufficient to describe it. 
Moreover, the charged Kerr solutions are of great interest since, as proved by Mazur 
\cite{Mazur1982}, they are the only stationary, axisymmetric electrovac (vacuum without 
including the electromagnetic field) solutions describing a black hole (see Ref. 
\cite{Carter1968} for a broader analysis). Moreover, there exist several examples 
\cite{Wald1972} on how the final black hole state resulting from the gravitational 
collapse can lose all information about the collapsing body with exception of its mass, 
angular momentum and charge. 
It would be an error to consider the aforementioned metrics as independent models. Indeed, 
they are all related and deeply connected to each other, belonging to the same class 
of solutions of the Einstein equations. The work 
in Ref. \cite{NJ1965} has even suggested that the Kerr metric can be formally obtained by 
applying (in a suitable way) a complex coordinate transformation $r \rightarrow r+ia\cos\theta$ 
to the Schwarzschild one. 

All black hole solutions are examples of Kerr-Schild metrics \cite{KS1965}, 
defined by imposing on the components of the metric $g$, solution of 
the Einstein equations, the structure
\begin{equation}
    g_{\mu\nu}=\eta_{\mu\nu}+\mathcal{F}l_\mu l_\nu,
    \label{kerr-Schild form of the metric}
\end{equation}
where $\eta$ is the Minkowski metric, $\mathcal{F}$ is a scalar function and $l$ is a 
vector field required to be null with respect to $g$ (and hence with respect to $\eta$ as well). 
Another interesting result is that such metrics can be endowed with electric charge without 
modifying their associated null congruence \cite{DKS1969}.

After such a brief classification of the Kerr-Newman solution in the 
context of black hole physics, we want to highlight some features of 
this space-time which will be useful for our purposes. Among the several 
coordinate systems in which one can write the metric, we would like to 
mention the Kerr-Schild one (cf. Ref. \cite{AN2014} for a review): in 
this way the metric is explicitly written in the Kerr-Schild form 
\eqref{kerr-Schild form of the metric}, where the Minkowski metric 
appears in oblate spheroidal coordinates. This aspect will propagate 
throughout computations, giving us spheroidal wave functions governing 
the angular part of the scalar field we will study. A coordinate system 
which minimizes the number of off-diagonal metric components (and is 
therefore well suited for explicit computations) is the Boyer-Lindquist 
one \cite{BL1967}. Other advantages are the presence of an explicit 
timelike coordinate and the evident asymptotic flatness. Upon considering 
a metric with signature $-2$ and introducing standard spherical coordinates 
($t,r,\theta,\phi)$, where $t\in [-\infty,\infty]$, $r\ge0$, $\theta\in[0,\pi]$, 
$\phi\in[0,2\pi[$, with $G=c=1$ units, the metric in BL coordinates reads as
\begin{equation}
\begin{split}
ds^2=&\left(\frac{\Delta  - a^2\sin^2\theta}{\rho^2}\right)dt^2
+\frac{2a(r^2+a^2-\Delta)\sin^2\theta}{\rho^2}dtd\phi \\& -\left[\frac{(r^2+a^2)^2
-\Delta a^2\sin^2\theta}{\rho^2}\right]\sin^2\theta d\phi^2
+\frac{\rho^2}{\Delta}dr^2+\rho^2d\theta^2,
\label{kerr-newman metric in bl coordinates}
\end{split}
\end{equation}
where 
\begin{equation}
    \rho^2=r^2+a^2\cos^2\theta,
    \label{rhoquadro kerr-newman bl coordinates}
\end{equation}
\begin{equation}
    \Delta=r^2+a^2+Q^2-2Mr.
    \label{Delta for the kerr newman metric}
\end{equation}
As stated before, this stationary, axisymmetric (for these reasons metric 
components are $t,\phi$ independent) and asymptotically flat solution of 
the Einstein-Maxwell equations describes the space-time of an isolated 
black hole. It is easy to notice the invariance of the metric under the 
transformation $\{t,\phi\}\rightarrow \{-t,-\phi\}$, i.e., reversing the 
time and the rotation direction of the black hole leaves the space-time 
unaffected. The three parameters $M,Q,a$ have a direct physical interpretation: 
$M$ is the total mass of the black hole while $ a=J/M$ is the angular 
momentum per unit mass of the black hole as measured by a distant observer. 
For the parameter $Q$ there exist instead 
different definitions in the literature: 
it is sometimes referred to as the total electric charge of the black hole, 
other times the total electric charge of the whole space-time. Upon introducing 
the natural volume element $\epsilon=\sqrt{-g} dt\wedge dr\wedge d\theta\wedge 
d\phi$ of the manifold and the Faraday-Maxwell tensor $\mathbf{F}$, for every 
$2$-sphere $S$ in the asymptotic region we have 
\begin{equation}
    \frac{1}{2}\int_S\mathbf{\epsilon} \cdot \mathbf{F}=4\pi Q.
\end{equation}
Henceforth, as suggested by Wald \cite{Wald1984}, $Q$ can be viewed 
as the total electric charge of the space-time.
The connection among the black hole metrics can be easily seen also from 
an analytical perspective: in the $Q\rightarrow0$ limit one obtains the 
vacuum Kerr solution, for $a\rightarrow0$ the Reissner-Nordstrom one and for 
$\{Q\rightarrow0, a\rightarrow0\}$ the Schwarzschild solution.

The metric exhibits two singularities (the same of Kerr), which can 
be read directly from its expression. One of them is a physical singularity 
(scalars from the Riemann tensor diverge here), given by the condition
\begin{equation}
\rho^2=0.
\end{equation}
This equation is satisfied by the values $\{r=0, \theta=\pi/2\}$. However, 
in this coordinate system, its physical meaning is hard to reveal. In order 
to give a reasonable interpretation of it, it is better to consider quasi-Cartesian 
coordinates \cite{KS2009}, where it translates into the conditions $\{z=0, 
x^2+y^2=a^2\}$. For this reason, it takes the name of \textit{ring} singularity, 
splitting the whole space-time into two unattached regions. In order to deal with 
it, one can consider two copies of the Kerr-Newman space-time, remove the ring 
and then attach such copies by identifying their sides. An exhaustive description 
of such a topological procedure can be found in Ref. \cite{HE1973}. Another 
insight is given by considering the case of vanishing gravitational constant: 
we have a Maxwell field in Minkowski space-time sourced by a rotating charged 
disk with boundary \cite{LYNDEN2004, KAISER2004}. It is worth mentioning that 
$r$ might be analytically continued also to negative values through such a disk. 
However, this region is not of physical interest (at least for our purposes): 
for instance, at large values of $r$ we would end up again in an asymptotically 
flat space-time but this time with a negative mass. We can therefore, without 
loss of generality, restrict ourselves to the usual $r>0$ interval. 
The other singularity, 
 \begin{equation}
     \Delta=(r-r_+)(r-r_-)=0,
 \end{equation}
is a pseudo-singularity depending on the coordinate system we have considered. 
However, even if it is a coordinate artifact, it has an interesting geometrical 
description, in the same spirit of the $r=2M$ singularity in the Schwarzschild 
case. The two points singled out,
\begin{equation}
r_{\pm}=M\pm\sqrt{M^2-a^2-Q^2},
\label{two horizons of the kerr newman metric}
\end{equation}
represent the two horizons of the black hole: the inner (or Cauchy) horizon 
$r_-$ and the outer (event) horizon $r_+$. In fact, $\Delta=0$ surfaces are 
null hypersurfaces, since the normal vector to r-constant surfaces satisfies 
$n_\alpha n_\beta g^{\alpha \beta}=g^{rr}=\Delta/\rho^2$. 
We have therefore three regions of interest (cf. Ref. \cite{HE1973} for 
Penrose-Carter diagrams and more details):
\begin{itemize}
    \item Region I: $0<r<r_-$. It contains the ring singularity. Furthermore, 
there exist closed timelike curves through every point, violating causality.
    \item Region II: $r_-<r<r_+$. Here, the surfaces of constant $r$ are 
spacelike. Moreover, there exist closed trapped surfaces. Hence, an observer 
crossing the $r=r_-$ surface would see the whole of the history of the 
region outside the black hole in a finite amount of time. Objects in this 
region appear infinitely blue-shifted as they approach future spacelike 
infinity. Moreover, the $r_-$ surface shows also instability problems in 
the initial data formulation of general relativity.  
This feature is already present in the 
Reissner-Nordstrom and Kerr space-time.
    \item Region III: $r>r_+$. It represents jut the asymptotically 
flat region, the exterior of the black hole. 
\end{itemize}
The uncharged Kerr version has a very similar global structure.

The points $r_{\pm}$ give us also a constraint between the parameters: 
since $r$ must be real (there is no need to rely on a complex space-time), 
the physical scenario is given by the region of the parameter space where
\begin{equation}
a^2+Q^2\le M^2.
\end{equation}
When the equality holds, the two horizons collapse into one and this scenario 
is called the extreme Kerr-Newman black hole: it has a slightly different 
causal structure since in this case the region II does not exist. In the 
other case, i.e. $a^2+Q^2>M^2$, the ring singularity is the only one, without 
an event horizon. We therefore end up with only a naked singularity which 
violates the cosmic censor conjecture, leading to a metric which does not 
describe a black hole. Moreover, it is worth highlighting that in several 
astrophysically reasonable situations, it appears that $Q\ll M$, giving us 
in principle the possibility to rely on a series expansion with respect to 
$Q$ for more involved computations.
Even if not strictly related to the causal structure but rather to the motion 
of particles in the space-time, the last region of interest is the so-called 
ergosphere. Its range is defined by $r_+<r<r_+^E$, where 
\begin{equation}
r_+^E(\theta)=M+\sqrt{M^2-a^2\cos^2\theta -Q^2}.
\end{equation}
This results from the vanishing of the $g_{tt}$ component of the metric. 
Its physical meaning is strictly connected to the Mach idea of the dragging 
of inertial frames: in such a region, a particle rotates in the direction 
given by the black hole, leading to the impossibility of having a particle 
at rest with respect to a distant observer (cf. Ref. \cite{AN2014} for 
more details regarding the last reported features).

The Kerr-Newman metric is, however, only the gravitational part of a solution 
to the Einstein-Maxwell equations. The associated Maxwell field is given 
by the electromagnetic $1$-form having components
\begin{equation}
A_\mu=\frac{Qr}{\rho^2}(1,0,0,-a\sin^2\theta).
\end{equation}

\subsection{Interlude: why can we separate variables?}

The special algebraic type of the space-time, complete integrability of 
geodesic motion, and separability of the Hamilton-Jacobi, Klein-Gordon and 
Dirac equations are all properties which, at a deeper level, are strictly 
connected to each other. Their common root is the so-called {\it principal tensor}, 
a tensor which can be seen as a ``seed object": one can generate all possible 
symmetries that a space-time admits; furthermore, it determines uniquely the form 
of the geometry. By following Ref. \cite{Frolov2017kze}, we would like to 
highlight how black hole metrics are well-suited for the application of 
our method, since they obey field equations which are separable.  
The concept of hidden symmetries seems to be a good starting point. 
\newline
A space-time has a symmetry if there exists a transformation which 
preserves its geometry. This means that the metric, as well as all other 
quantities constructed from it, remains unchanged by such a transformation. 
Continuous symmetry transformations are generated by Killing vector fields; 
we call the corresponding symmetries explicit. By Noether's theorem, they 
generate conserved charges. Killing tensors are in one-to-one correspondence 
with constants of geodesic motion that are homogeneous in particle momenta, 
i.e., a rank $r$ Killing tensor gives rise to a homogeneous constant of 
motion of degree $r$ in momentum. The best known example of a Killing 
tensor is the space-time metric itself. The corresponding conserved quantity 
is the Hamiltonian for the relativistic particle and its value is proportional 
to the square of particle's mass. The geometric structure of space-time 
encoded in Killing tensors of rank $2$ and higher is called a hidden symmetry. 
If the space-time has some symmetries they can be always ``lifted up" to the 
phase space (motion of particles on it) symmetries. The contrary is not true. 
Symmetries which have the direct counterpart on the configuration space will 
be called explicit, those which cannot be reduced to configuration space 
transformations are called hidden symmetries.
We are thus considering not only the space-time itself, but also the dynamics 
on it in order to understand the relationship among the motion of particles 
and/or fields and the underlying geometry.

\subsection{Kerr-NUT-A(d)S black holes}

The existence of the principal tensor imposes very restrictive conditions 
on the geometry. In fact, it determines the geometry: the most general consistent 
with the existence of the principal tensor is the off-shell Kerr-NUT-(A)dS geometry. 
In order to pave the way to this metric, following slightly the notation of 
Ref. \cite{Frolov2017kze}, we start by considering the Kerr metric in 
Boyer-Lindquist coordinates (which can be obtained from the Kerr-Newman one 
given in Eq. \eqref{kerr-newman metric in bl coordinates} by setting $Q=0$)
\begin{equation}
    ds^2=-\left(1-\frac{2Mr}{\rho^2}\right)dt^2-\frac{4Mra\sin^2\theta}
{\rho^2}dtd\phi+\frac{\mathcal{A} \sin^2\theta}{\Delta_r}dr^2+\rho^2 d\theta^2,
    \label{kerr metric in bl coordinates}
\end{equation}
where $\rho^2$ is given by Eq. \eqref{rhoquadro kerr-newman bl coordinates}, 
$\Delta\equiv\Delta_r$ by Eq. \eqref{Delta for the kerr newman metric} 
(setting $Q=0$) and $ \mathcal{A}=(r^2+a^2)^2-\Delta_ra^2\sin^2\theta$.

The metric does not depend on coordinates $t$ and $\phi$ (stationary and 
axial symmetry), therefore
\begin{equation}
\xi_{(t)}=\partial_t, \quad \xi_{(\phi)}=\partial_\phi,
\end{equation}
are the two (commuting) Killing vectors. Furthermore, these are the only 
Killing vectors that the space-time admits. However, the scenario is much wider 
when one studies not only the space-time itself, but also the dynamics of 
fields on it. Note how poor are space-time symmetries and how rich will be 
the set of hidden symmetries. In order to study them, it is useful to introduce 
the so-called canonical coordinates (or Carter's coordinates)
\begin{equation}
    y=a\cos\theta, \quad \psi=\frac{\phi}{a}, \quad \tau =t-a\phi.
\end{equation}
The Kerr metric takes therefore the form
\begin{equation}
    ds^2=-\frac{\Delta_r}{\rho^2}(d\tau+y^2d\psi)^2+\frac{\Delta_y}
{\rho^2}(d\tau-r^2d\psi)^2+\frac{\rho^2}{\Delta_r}dr^2+\frac{\rho^2}{\Delta_y}dy^2,
    \label{off shell canonical kerr nut ads metric}
\end{equation}
where now  $\rho^2=r^2+y^2$.
The functions $\Delta_r=\Delta_r(r)$ and $\Delta_y=\Delta_y(y)$ contain the 
physical parameters $M,a$. If they are not specified, the metric is said to 
be off-shell. This is useful to handle general computations without relying 
on a specific black-hole model. Upon introducing the cosmological constant 
$\Lambda$ and the NUT parameter $N$ throughout $\Delta_r(r)$ and $\Delta_y(y)$, i.e.,
\begin{equation}
    \begin{split}
        \Delta_r=&(r^2-a^2)\left(1-\Lambda\frac{r^2}{3}\right)-2Mr,\\
        \Delta_y=&(a^2-y^2)\left(1+\Lambda\frac{y^2}{3}\right)+2Ny,
    \end{split}
\end{equation}
we cover the Kerr-NUT-(A)dS family of space-times. By setting $\Lambda=0$ and 
$N=0$ we recover the Kerr one. In addition to $M$ and $a$, we have the 
cosmological constant $\Lambda$ and the NUT parameter $N$.
It is possible to achieve a more symmetric form by writing 
$x=ir, b_x=iM, b_y=N$, hence obtaining
\begin{equation}
    \begin{split}
        \Delta_x=&(a^2-x^2)\left(1+\Lambda\frac{x^2}{3}\right)+2b_xx,\\
        \Delta_y=&(a^2-y^2)\left(1+\Lambda\frac{y^2}{3}\right)+2b_yy,
    \end{split}
\end{equation}
and
\begin{equation}
    ds^2=\frac{\Delta_y}{(y^2-x^2)}(d\tau+x^2d\psi)^2+\frac{\Delta_x}
{(x^2-y^2)}(d\tau+y^2d\psi)^2+\frac{(y^2-x^2)}{\Delta_y}dy^2
+\frac{(x^2-y^2)}{\Delta_x}dx^2,
\end{equation}
which is now the metric in a symmetric form with respect to the formal 
substitution $x\leftrightarrow y$. 

\subsection{Pleba\'nski-Demia\'nski black holes}

We could, in principle, further generalize the Kerr-NUT-(A)dS metric by 
adding electric and magnetic charge and an acceleration parameter to the 
black-hole. The Pleba\'nski-Demianski metric is the most general four-dimensional 
electrovacuum solution of Einstein equations that is stationary, axisymmetric, 
and whose Weyl tensor is of the special algebraic type Petrov D. In the 
off-shell metric given in Eq. \eqref{off shell canonical kerr nut ads metric}, we must 
multiply it by an overall factor $\Omega^2=(1-yr)^{-2}$ and
\begin{equation}
\begin{split}
\Delta_r=&k+e^2+g^2-2mr+\varepsilon r^2-2nr^2-\left(k+\frac{\Lambda}{3}\right)r^4,\\
\Delta_y=&k+2ny-\varepsilon y^2+2my^3-\left(k+e^2+g^2+\frac{\Lambda}{3}\right)y^4,
\end{split}
\end{equation}
where constants $k,m,\varepsilon,n$ are free parameters related to mass, rotation, 
NUT parameter and acceleration.
The electromagnetic side is given by the Faraday-Maxwell 2-form 
$\mathbf{F}=d\mathbf{A}$, where
\begin{equation}
    \mathbf{A}=-\frac{er}{\rho^2}(\mathbf{d}\tau +y^2\mathbf{d}\psi)
-\frac{gy}{\rho^2}(\mathbf{d}\tau-r^2\mathbf{d}\psi).
\end{equation}

\section{Massive scalar field in Kerr-Newman space-time}
\label{A massive scalar field in Kerr-Newman space-time}

Despite the absence of spherical symmetry, the Klein-Gordon equation still 
admits a solution via the separation of coordinates 
method in Kerr-Newman space-times. 
The separability of the $t$ and $\phi$ coordinates arises from the existence 
of the two Killing vectors $\partial_t, \partial_\phi$ by virtue of stationarity 
and axial symmetry of the metric. The possibility to separate also the other 
two coordinates arises from a set of striking features of the Kerr metric which are 
in part summarised in Ref. \cite{Teukolsky2015} and, as stated in Sec. 
\ref{Stationary, axial-symmetric black holes}, are deeply related to the existence 
of the principal tensor. The first who noticed the separability of the scalar 
wave equation in the Kerr metric was Carter \cite{Carter1968}. However, there is 
a relevant difference with respect to the Schwarzschild case that we would like 
to stress: the radial and angular equations are separated by an analytical function 
rather than a simple constant. Obtaining its exact value includes solving involved 
continued fractions. For explicit computations one relies instead on its asymptotic 
values, well known in the mathematical literature on spheroidal wave functions. 
Scalar fields display striking features in the Kerr-Newman space-time. The nonvanishing 
angular momentum of the black hole admits the presence of bound states around 
the event horizon, forming the so-called scalar clouds \cite{BCHR2014}. Moreover, 
the reflected part of an incident wave to the event horizon might have an amplitude 
greater than the former: this is called super-radiance (cf. Ref. \cite{FN2004} 
and related papers). It all depends on the relation among the frequency of the 
field and the critical frequency of the black hole.  
Super-radiance is studied by matching local solutions of the Klein-Gordon equation, 
the one at the event horizon with the one at infinity.

\subsection{The full equation}

The dynamics of a charged massive scalar field $\psi$, of mass $\mu$ and 
electric charge $e$, is given by the Klein-Gordon equation
\begin{equation}
    (\Box+\mu^2)\Psi(t,r,\theta,\phi)=0,
\end{equation}
which after considering the minimal coupling and the scalar nature of the field reads as
\begin{eqnarray}
&\left[\frac{1}{\sqrt{-g}}\partial_\mu(g^{\mu\nu}\sqrt{-g}\partial_\nu)
+ie(\partial_\mu A^\mu)+2ieA^\mu \partial_\mu  \right. \nonumber \\
&+\left.\frac{ie}{\sqrt{-g}}A^\mu(\partial_\mu\sqrt{-g})-e^2A^\mu 
A_\mu+\mu^2\right]\psi(t,r,\theta,\phi)=0.
\end{eqnarray}
We here rewrite the Kerr-Newman metric for convenience:
\begin{equation}
\begin{split}
ds^2=&\left(\frac{\Delta  - a^2\sin^2\theta}{\rho^2}\right)dt^2
+\frac{2a(r^2+a^2-\Delta)\sin^2\theta}{\rho^2}dtd\phi \\& -\left[\frac{(r^2+a^2)^2
-\Delta a^2\sin^2\theta}{\rho^2}\right]\sin^2\theta d\phi^2
+\frac{\rho^2}{\Delta}dr^2+\rho^2d\theta^2,
\end{split}
\end{equation}
where (cf. Eqs \eqref{rhoquadro kerr-newman bl coordinates}, 
\eqref{Delta for the kerr newman metric})
\begin{eqnarray}
     \rho^2=&&r^2+a^2\cos^2\theta,\\
       \Delta=&&r^2+a^2+Q^2-2Mr.
\end{eqnarray}

The metric components are $\{t,\phi\}$-independent since space-time is stationary 
and axisymmetric. Henceforth, it is still possible to rely on separation 
of variables, considering a Fourier series with respect to the angular coordinate 
$\phi$ and a Fourier integral with respect to the time $t$. Given an integer $m$, writing
\begin{equation}
    \psi(t,r,\theta,\phi)=\int_{-\infty}^{\infty} \frac{d\omega}{\sqrt{2\pi}}
e^{-i\omega t}\sum_{m=0}^{\infty} e^{im\phi} \sum_{n=m}^{\infty}  R^m_n(r)P^m_n(\theta),
    \label{coordinate decomposition}
\end{equation}
we obtain with fixed $m,n$ an ordinary differential equation for the radial 
$R^m_n(r)$ and angular $P^m_n(\theta)$ parts. We start by considering the latter, 
since it gives us the possibility to discuss the role of $m$ and $n$, which arise 
from the separation parameter $\lambda$, which (as stated in the introduction) by virtue 
of lack of spherical symmetry of space-time is no longer constant (in the 
spherical case it is just $l(l+1)$), but is an analytical function that we will denote 
by $\lambda^m_n$. From now on, we will analyse the case of a single mode for fixed $\omega$.

\subsection{The angular solution}

By virtue of the separation of variables, the angular equation reads
\begin{equation}
\left[\frac{1}{\sin\theta}\frac{d}{d\theta}\left(\sin\theta\frac{d}{d\theta}\right)
+\left(\lambda^m_n+a^2(\mu^2-\omega^2)\sin^2\theta
-\frac{m^2}{\sin^2\theta}\right)\right]P^m_n(\theta)=0.
\label{original angular equation}
\end{equation}
By exploiting the change of variables (cf. Refs \cite{BE1953,NIST}) 
\begin{equation}
    \gamma^2=a^2(\mu^2-\omega^2), \quad z=\cos\theta,
\end{equation}
we are able to write Eq. \eqref{original angular equation} in the well-known form
\begin{equation}
   \left[ (1-z^2)\frac{d^2}{dz^2}-2z\frac{d}{dz}+\left(\lambda^m_n(\gamma^2) 
+\gamma^2(1-z^2)-\frac{m^2}{(1-z^2)}\right)\right]P^m_n(z)=0,
    \label{fancy angular equation}
\end{equation}
which is a spheroidal differential equation. The three real parameters of interest 
are $\lambda^m_n(\gamma^2)$, $\gamma^2$ and $m$. In applications involving prolate 
spheroidal coordinates $\gamma^2$ is positive, while it is negative in applications 
involving oblate spheroidal coordinates. Eq. \eqref{fancy angular equation} exhibits 
two regular singularities at $z=\pm1$ (with exponents $\pm\frac{1}{2}m$) and one 
irregular singularity of rank $1$ at $z=\infty$ (if $\gamma\neq0$). However, we are 
interested in the $z\in[-1,1]$ interval (since $\theta\in [0,\pi]$). Since in our 
case $m$ is, by construction, a non-negative integer, Eq. \eqref{fancy angular equation} 
takes a simpler form, named spheroidal wave equation. 
The eigenvalues $\lambda^m_n=\lambda^m_n(\gamma^2)$ exhibit what is called the 
overtone number $n$, where $n=m,m+1,m+2 \dots$. It is useful to bear in mind 
some properties. We have:
\begin{equation}
    \lambda^m_m(\gamma^2)<\lambda^m_{m+1}(\gamma^2)<\lambda^m_{m+2}(\gamma^2)<...,
\end{equation}
\begin{equation}
    \lambda^m_m(\gamma^2) \rightarrow n(n+1)-\frac{1}{2}\gamma^2+O(n^{-2})\quad  
\text{when} \quad n\rightarrow \infty,
    \label{lambda large overtone number}
\end{equation}
\begin{equation}
    \lambda^m_n(0)=n(n+1).
\end{equation}
There also exist series expansions for large and small value of $\gamma$, i.e., 
for $\omega \simeq \mu$ and $\omega\gg\mu$ (fixed $a$). In fact,
\begin{equation}
    \lambda^m_n(\gamma^2) \rightarrow -4\gamma^2 +2\gamma(2n-2m+1)+O(1) 
\quad \text{for} \quad \gamma^2 \rightarrow \infty,
\end{equation}
while for small $\gamma^2$ we have
\begin{equation}
    \lambda^m_n(\gamma^2)\simeq n(n+1)-\frac{1}{2}\left(1+\frac{(2m-1)(2m+1)}{(2n-1)(2n+3)}
\right)\gamma^2 + {\rm O}(\gamma^{4}).
    \label{lambda small values of gamma2}
\end{equation}
As stated in the references, the power series expansion, up to the fifth power of 
$\gamma$ (inclusive), gives still quite useful approximations 
up to $|\gamma^2|=5$, approximately. 

The most general solution, which is bounded along the $z$-axis, is given in terms of 
the spheroidal functions of the first kind $Ps^m_n(z,\gamma^2)$ and second kind 
$Qs^m_n(z,\gamma)$, as follows:
\begin{equation}
    P(z)=P^m_n(z,\gamma^2)=c_1Ps^m_n(z,\gamma^2)+c_2Qs^m_n(z,\gamma^2),
    \label{general solution angular part}
\end{equation}
where $c_1$ and $c_2$ are integration constants. Depending on the sign of 
$\gamma^2$, they are prolate angular spheroidal wave functions for $\gamma^2>0$ 
and oblate angular spheroidal wave function for $\gamma^2=0$. For $\gamma^2=0$ 
Eq. \eqref{fancy angular equation} reduces to the Legendre differential equation and  
\begin{equation}
    P^m_n(z,0)=c_1P^m_n(z)+c_2Q^m_n(z),
    \label{general solution angular part}
\end{equation}
where $P^m_n(z)$ and $Q^m_n(z)$ are the Legendre functions of the first and 
second kind, respectively. This has a direct physical meaning since, aside 
the case $\mu^2=\omega^2$, $\gamma^2=0\implies a=0$, and hence the spherical 
symmetry of space-time is recovered (as in the Schwarzschild or Reissner-Nordstrom black hole). 
Furthermore, one can express $P^m_n(z,\gamma^2)$ as a series 
of Legendre functions, where the coefficients of this expansion can be easily 
found in the mathematical literature.  
We note, incidentally, that spheroidal wave functions can be seen as special 
cases of Heun functions, as clarified in Ref. \cite{VBM2014}. 

\subsection{Radial solution at regular singular points}

In light of the above, the radial equation is given by
\begin{equation}
\left[\frac{d^2}{dr^2}+p(r)\frac{d}{dr} +q(r)\right]R^m_n(r)=0,
\label{radial equation}
\end{equation}
where
\begin{equation}
    \begin{split}
        p(r)=&\left(\frac{1}{(r-r_+)}+\frac{1}{(r-r_-)}\right),\\
        q(r)=&\frac{[\omega(r^2+a^2)-am-eQr]^2}{(r-r_+)^2(r-r_-)^2}
-\frac{ [\lambda^m_n(\gamma^2)+\mu^2(r^2+a^2)-2am\omega]}{(r-r_+)(r-r_-)}.
    \end{split}
\end{equation}
It is easy to see that such an ordinary differential equation shows two regular 
singular points of rank $2$ at $r=r_\pm$ and an irregular singular point of rank 
$1$ at $r\rightarrow \infty$. For this reason, it belongs to the set of Heun 
equations, in particular it is a confluent Heun equation in the so-called non 
symmetrical canonical form \cite{Ronveaux1995}. These types of equations are studied 
by defining the variable (standard step in the mathematical literature)
\begin{equation}
    x=\frac{(r-r_-)}{(r_+-r_-)}.
\end{equation}
In fact, such a transformation maps the two regular singularities $\{r_-,r_+\}$ 
to $\{0,1\}$, leaving unchanged the one at infinity. In this way the equation 
takes the simplest form, allowing an easy reading of the parameters. A simple 
overview of this topic is made in Ref. \cite{ER}. Here, we follow the detailed 
analysis given in Ref. \cite{VBM2022} in order to write the local solutions 
$R_0^{mn}(x), R_1^{mn}(x)$ at $x=0,1$ respectively. The choice to exploit the 
$x$ coordinate is just to make the notation easier. Upon defining the non-symmetrical 
canonical factor as
\begin{equation}
    \xi(x;\alpha,\beta,\gamma)=x^{\beta/2}(x-1)^{\gamma/2}e^{\alpha x/2},
\end{equation}
we have
\begin{equation}
\begin{split}
    R_{0}^{mn}(x) =& \xi(x;\alpha,\beta,\gamma)
[c_1\text{HeunC}(\alpha,\beta,\gamma,\delta,\eta;x)+c_2x^{-\beta}
\text{HeunC}(\alpha,\beta,\gamma,\delta,\eta;x)], \\ 
R_{1}^{mn}(x) =& \xi(x;\alpha,\beta,\gamma) [c_{1}
{\rm HeunC}(-\alpha,\gamma,\beta,-\delta,\eta+\delta;-x)\\&
+ c_{2}(x-1)^{-\gamma}{\rm HeunC}(-\alpha,-\gamma,\beta,-\delta,\eta + \delta;x)],
\end{split}
\end{equation}
where $\text{HeunC}(\alpha,\beta,\gamma,\delta,\eta;x)$ is the confluent Heun 
function \cite{Ronveaux1995} and $c_1,c_2$ integration constants. The parameters 
are instead given by \cite{VBM2022}
\begin{equation}
    \begin{split}
        \alpha=&\pm2(r_+-r_-)\sqrt{\mu^2-\omega^2},\\
        \beta=&\pm2i\frac{[\omega(r^2_-+a^2)-a m-eQr_{-}]}{(r_+-r_-)},\\
        \gamma=&\pm2i\frac{[\omega(r^2_++a^2)-am-eQr_{+}]}{(r_+-r_-)},\\
        \delta=&-(r_+-r_-)[2eQ\omega+(r_++r_-)(\mu^2-2\omega^2)],\\
        \eta=&-\frac{1}{(r_+-r_-)^2}\{a^2[2m^2+\mu^2(r_+-r_-)^2]
+2amQe(r_++r_-)\\&+2Q^2r_+r_-e^2+(r_+-r_-)^2(\lambda^m_n+\mu^2r_-^2)\\&
-2\omega(2a^3m+a^2Qr_+e+a^2Qr_-e+amr^2_++amr_-^2+eQr_+r^2_-e-Qr^3_-e)\\&
+2\omega^2(a^2+r_-^2)(a^2+2r_+r_--r_-^2)\}.
    \end{split}
\end{equation}
The leading terms of the two local solutions are given by
\begin{eqnarray}
R_{0}^{mn}(x) &\sim&  x^{\beta/2}(x-1)^{\gamma/2}
e^{\alpha x/2}(c_{1}+c_{2}x^{-\beta}),  \\ \nonumber \\
     R_{1}^{mn}(x)& \sim& x^{\beta/2}(x-1)^{\gamma/2}
e^{\alpha x/2}[c_{1}+c_{2}(x-1)^{-\gamma}],
\end{eqnarray}
respectively. The sign choice of $\alpha,\beta,\gamma$ 
depends on the particular phenomenon one 
wants to study. For instance, choosing the negative sign leads to ingoing waves 
at the outer event horizon, as can be seen from the leading behaviour of 
$R_1(x)$. In conclusion, we report also the asymptotic solution 
of Eq. \eqref{radial equation},
\begin{equation}
    R_\infty^{mn}(x)\sim x^{\beta/2}(x-1)^{\gamma/2}
e^{\alpha x/2}x^{-(\frac{\beta+\gamma+2}{2}
+\frac{\delta}{\alpha})}(c_1+c_2e^{-\alpha x}).
\end{equation}
The mathematical theory is not able to give an exact solution for this region, 
but relies on asymptotic expansions due principally to Thom\'e.

\subsection{Radial solution at spacelike infinity}

Heun's theory provides local solutions in the neighbourhood of singular points. 
In principle one could extend via analytical continuation such solutions in order 
to cover broader regions of the space considered. An alternative, as already stated 
in Sec. \ref{Poincare asymptotic series expansion}, is given by the asymptotic 
series expansion developed by Poincar\'e: for linear differential equations with 
variable coefficients (as Eq. \eqref{radial equation}), the solution consists of 
the limiting form at the origin times the limiting form at infinity times a suitable 
series which interpolates between these two cases. 
In our models we are dealing with more than 
one regular singular point (we have two horizons), adding another piece to the whole 
series solution. In order to construct such a function, we have to reduce Eq. 
\eqref{radial equation} to its canonical form. In the literature (cf., e.g., Ref. 
\cite{BCHR2014}) this is achieved upon considering the tortoise 
coordinate $r_*$, defined by the relation
\begin{equation}
    \frac{dr_*}{dr}\equiv \frac{(r^2+a^2)}{\Delta}.
\end{equation}
The differential equation is then written in a mixed way: the derivatives are taken 
with respect to $r_*$ while the resulting potential is written in terms of the standard 
radial coordinate $r$. However, there is no need to introduce 
other coordinate systems, since one can consider the Laguerre-Forsyth \cite{Wilc} 
canonical reduction \cite{Esposito2017}. Here, we rewrite the 
formulae already displayed in Sec. \ref{Poincare asymptotic series expansion} in order 
to adapt the notation (especially indices) to this particular case. We start from the 
ansatz that the radial solution $R^m_n(r)$ can be factorized in the following way  
(no confusion should arise with the coefficients defined in the previous section):
\begin{equation}
    R^m_n(r)=\alpha(r)\beta^m_n(r),
    \label{decomposition ansatz radial solution canonical}
\end{equation}
By inserting such a decomposition into the original radial equation 
\eqref{radial equation}, one finds that
\begin{equation}
\alpha(r)=\exp\left(-\frac{1}{2}\int dr p(r)\right)
= \frac{1}{\sqrt{r-r_+}}\frac{1}{\sqrt{r-r_-}},
\label{alpha part radial canonical solution}
\end{equation}
while $\beta^m_n(r)$ is the solution of the canonical differential equation
\begin{equation}
    \left[\frac{d^2}{dr^2}+J^m_n(r)\right]\beta^m_n(r)=0.
    \label{canonical equation}
\end{equation}
The canonical potential $J^m_n(r)$ is given by (cf. Ref. \cite{ER})
\begin{equation}
    \begin{split}
        J^m_n(r)=&-\frac{1}{2}\frac{d}{dr}p(r)-\frac{1}{4}p^{2}(r)+q(r)\\
        =&\frac{1}{2}\left(\frac{1}{(r-r_+)^2}+\frac{1}{(r-r_-)^2}\right)
-\frac{1}{4}\left(\frac{1}{(r-r_+)}+\frac{1}{(r-r_-)}\right)^2\\
        &+\frac{1}{\Delta^2}\{[\omega(a^2+r^2)-am-eQr]^2
-\Delta[\lambda^m_n(\gamma^2)+\mu^2(a^2+r^2)-2am\omega]\}.
    \end{split}
\end{equation}
By direct inspection, $r_\pm$ are regular singular points while the irregularity 
of the singular point at $r\rightarrow\infty$ can be easily seen by considering 
the change of variable $\rho=1/r$ and noting that $J^m_n(\rho)$ 
shows a fourth-order pole at $\rho=0$.
Note that in the Schwarzschild limit (i.e., $e,a,Q\rightarrow0$ and 
$\lambda^m_n(\gamma^2)\rightarrow l(l+1)$), such a potential coincides with the one we 
have recovered from the Schwarzschild radial equation in Ref. \cite{ER} (in the 
same paper we give more details about the following discussion).

The next step is to consider the following Poincar\'e asymptotic expansion for 
the potential $J^m_n(r)$ as r$\rightarrow \infty$:
\begin{equation}
     J^m_n(r)\approx \sum_{k=0}^\infty\frac{A_k^{mn}(\omega,\mu,e,a,Q,M)}{r^k},
     \label{canonical series expansion potential}
\end{equation}
where, for instance (we write only the $k$ sub-index to simplify the notation),
\begin{equation}
    \begin{split}
        A_0=&\omega^2-\mu^2, \label{A_0 massive ker newman}\\
        A_1=&2M(2\omega^2-\mu^2)-2e\omega Q,\\
        A_2=&(e^2+\mu^2-2\omega^2)Q^2-8MQe\omega-4M^2\mu^2
+12M^2\omega^2-\lambda^m_n(\gamma^2),\\
        A_3=&8M^3(4\omega^2-\mu^2)-24M^2Qe\omega +2M \Bigr[-2\omega^2(3Q^2+a^2)
-2am\omega\\&+2Q^2(e^2+\mu^2)+a^2\mu^2-\lambda^m_n(\gamma^2)\Bigr]
+2eQ\left[\omega(2Q^2+a^2)+am\right].
    \end{split}
\end{equation}
The black hole and field electric charge start to be relevant only for 
$k\ge1$, while the angular momentum of the black hole for $k\ge3$. At order zero, 
none of such features are encoded, leading just to plane waves propagating 
with frequency $\sqrt{\omega^2-\mu^2}$. With this decomposition in mind, the asymptotic 
form of Eq. \eqref{canonical equation} reads
\begin{equation}
    \left(\frac{d^{2}}{dr^{2}}+\sum_{k=0}^{\infty}
\frac{A_{k}^{mn}(\omega,\mu,e,a,Q)}{r^{k}}\right)\beta^m_n(r)=0.
\label{canonical series equation beta}
\end{equation}
If one considers only the $k=0,1,2$ terms, one obtains a solution at spacelike infinity 
in terms of Whittaker functions, as already pointed out in Ref. \cite{RW1977}. Within 
our framework, one can go beyond such an approximation by computing in principle every 
term of the series expansion given by Eq. \eqref{canonical series expansion potential}. 
The solution which results from Poincar\'e's theory has the form 
\begin{equation}
    \beta^m_n(r) \sim e^ {\varepsilon r \sqrt{-A_{0}}} \; r^\zeta
\; \left(1+\sum_{s=1}^{\infty}
\frac{B_{s}^{mn}(\omega,\mu,e,a,Q)}{r^{s}}\right).
\label{beta series solution}
\end{equation}
Here $\varepsilon=\pm1$, depending on the boundary condition one wants to impose at 
spacelike infinity. Since we are interested in solutions that approach zero 
far from the black hole, we impose $\varepsilon=-1$.  
From Eqs \eqref{decomposition ansatz radial solution canonical}, 
\eqref{alpha part radial canonical solution}, \eqref{beta series solution}, 
we obtain the full radial solution 
\begin{equation}
     R^m_n(r)=\mathcal{N}\frac{e^ {-r \sqrt{-A_{0}}} \; r^\zeta
\; }{\sqrt{(r-r_+)(r-r_-)}}\left(1+\sum_{s=1}^{\infty}
\frac{B_{s}^{mn}(\omega,\mu,e,a,Q)}{r^{s}}\right),
\label{our full radial solution}
\end{equation}
where $\mathcal{N}$ is an overall normalization constant.
The parameters $B_{s}^{mn}(\omega,\mu,e,a,Q)$, which form the interpolating 
series, can be determined by inserting Eq. \eqref{beta series solution} into 
Eq. \eqref{canonical series equation beta} and imposing the vanishing of the 
$r^{-s}$ coefficients. Since these conditions are independent, one obtains a 
linear system which can be solved easily via formal algebraic programs. 
Here, we report the exponent $\zeta$ of the power term, 
\begin{equation}
    \zeta=-\frac{[\mu^2M-2M\omega^2+e\omega Q]}{\sqrt{\mu^2-\omega^2}},
\end{equation}
and the first $B$ coefficient,
\begin{equation}
    \begin{split}
        B_1^{mn}(\omega,\mu,e,a,Q,M)=&\frac{1}{2(\mu^2-\omega^2)^2}
\biggr \{\sqrt{\mu^2-\omega^2}\Bigr[2\omega^4(4M^2-Q^2)-4MQe\omega^3
-\omega^2\lambda^m_n(\gamma^2) \\&
+3\omega^2(Q^2-4M^2)+6MQe\mu^2\omega \\&
+\mu^4(3M^2-Q^2)
+\mu^2(\lambda^m_n(\gamma^2)-Q^2e^2)\Bigr] \\&
-(\mu^2-\omega^2)(\mu^2M-2M\omega^2+e\omega Q)\biggr \}.
    \end{split}
\end{equation}

\subsection{The whole spectrum analysis at spacelike infinity for the s-wave}

The solution given by Eq. \eqref{our full radial solution} makes it possible for us to 
investigate the full frequency spectrum given by Eq. \eqref{coordinate decomposition}. 
We start by considering the fundamental mode $\{m,n\}=\{0,0\}$, i.e.
\begin{equation}
    \psi^0_0(t,r,\theta,\phi)=\int_{-\infty}^{\infty} 
\frac{d\omega}{\sqrt{2\pi}}e^{-i\omega t}R^0_0(r)P^0_0(z,\gamma^2). 
\end{equation}
An analytical investigation of such an integral, even if just in approximate form, 
can provide a picture of the behaviour of the full spectrum of the field, going 
beyond the single-Fourier mode analysis usually found in the literature. A first 
simplification arises since $\psi_0^0$ is independent of the $\phi$ coordinate. 
Moreover, since we want to focus our attention on the spacelike infinity region of 
space-time, we can freeze the $t$ coordinate to a certain arbitrary value and we can 
consider our asymptotic expansion in Eq. \eqref{our full radial solution}, in particular 
the leading order. By highlighting the overall phase $\varphi(r)$ of the exponential 
term, the previous integral takes a simpler form given by 
\begin{equation}
    \psi^0_0(t,r,\theta,\phi)=\frac{\mathcal{N}}{\sqrt{(r-r_+)(r-r_-)}}
\int_{-\infty}^{\infty} \frac{d\omega}{\sqrt{2\pi}}e^{-\varphi(r)} P^0_0(z,\gamma^2),
\label{full spectrum solution}
\end{equation}
where 
\begin{equation}
    \varphi(r)=i\omega t +r\sqrt{\mu^2-\omega^2}+\frac{[\mu^2 M
+e\omega Q-2M\omega^2]}{\sqrt{\mu^2-\omega^2}}\ln r ,
    \label{whole spectrum phase}
\end{equation}
and  $P^0_0(z,\gamma^2)$ is given by Eq. \eqref{general solution angular part} with 
$m=n=0$. Two remarks are in order: first, depending on the relationship among 
$\mu^2$ and $\omega^2$, either $\varphi(r)$ is purely imaginary 
or it has also a real part. Moreover, the explicit 
computation faces one important problem: the 
frequency dependence occurs not only in the radial term $e^{-\varphi(r)}$, but 
also in the angular one $P^0_0(z,\gamma^2)$ through $\gamma^2$. In order to disentangle 
this dependence and hence to obtain a frequency-independent radial part (the integral 
over the exponent is straightforward), one can expand the spheroidal wave solution, 
given by Eq. \eqref{general solution angular part}, in terms of Legendre functions, 
transferring the $\omega$-dependence to the coefficients of such an expansion. For 
these reasons, it is convenient to split the integral into two pieces, considering 
whenever $\sqrt{\mu^2-\omega^2}$ (and hence $\gamma$) is real or complex,
\begin{eqnarray}
     &&\psi^{0I}_0(t,r,\theta,\phi)=\frac{\mathcal{N}}{\sqrt{(r-r_+)(r-r_-)}}
\left(\int_{-\infty}^{-\mu}+\int_{\mu}^{\infty}\right) \frac{d\omega}{\sqrt{2\pi}}  
\Omega^I(r,z,\gamma^2), \label{full spectrum outside}\\
    &&\psi^{0II}_0(t,r,\theta,\phi)=\frac{\mathcal{N}}{\sqrt{(r-r_+)(r-r_-)}}
\int_{-\mu}^{\mu}\frac{d\omega}{\sqrt{2\pi}}  \Omega^{II}, \label{full spectrum inside}
\end{eqnarray}
where
\begin{eqnarray}
&&\Omega^I(r,z,\gamma^2)=e^{-\varphi^I(r)}P^0_0(z,\gamma^2), \label{omega1}\\
&&\varphi^I(r)=i\left(\omega t+r\sqrt{\omega^2-\mu^2}
+\frac{[\mu^2M+e\omega Q-2M\omega^2]}{\sqrt{\omega^2-\mu^2}}\ln r\right),
\end{eqnarray}
and
\begin{equation}
    \Omega^{II}=e^{-\varphi(r)} P^0_0(z,\gamma^2).
\end{equation}
This notation makes it possible for us to focus first on the functions 
$\Omega^{I,II}(r,z,\gamma^2)$ and then on the whole integral. 

\begin{figure}
    \centering
    \includegraphics[width=0.75\linewidth]{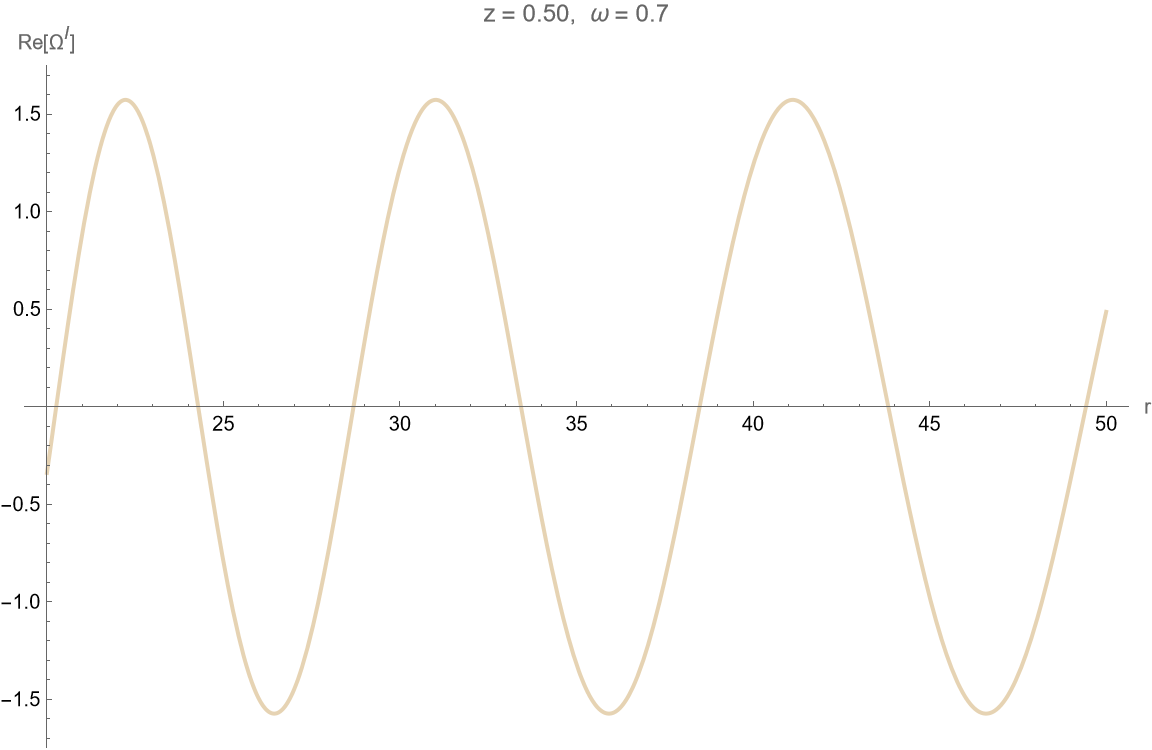}
    \caption{Real part of $\Omega^I(r,z,\gamma^2)$, given by Eq. \eqref{omega1}, 
plotted in the interval $r\in[20,60]$ for the following choice of parameters: 
$M=6, Q=1, a=1.1, \mu=0.6,e=0.5,t=1.$ 
The parameters $z$ and $\omega$ are instead considered as 
variable, in this case $z=0.5, \omega=0.7$. The expected behaviour is just the one 
of an oscillating wave.}
    \label{fig1}
\end{figure}

\begin{figure}
    \centering
    \includegraphics[width=0.75\linewidth]{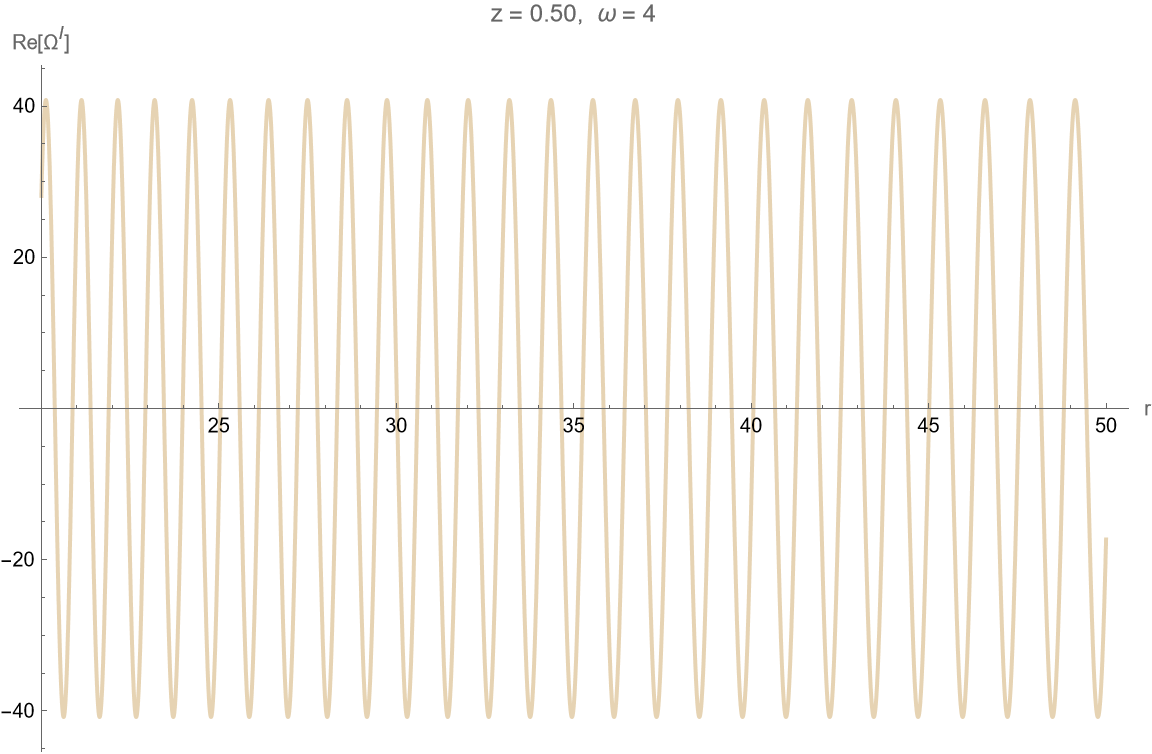}
    \caption{Real part of $\Omega^I(r,z,\gamma^2)$, given by Eq. \eqref{omega1}, 
plotted in the interval $r\in[20,60]$. The parameters are the same as in Fig. \ref{fig1}, 
but this time $\omega=4$. As $\omega$ increases, the wave exhibits stronger oscillations. 
Such a behaviour does not change for different values of the other parameters or for 
$z$. Negative values of $z$ and $\omega$ lead to the same qualitative trend.}
    \label{fig2}
\end{figure}

As we have stressed, in Eq. \eqref{full spectrum outside} the radial part is a 
purely imaginary exponent. In Eq. \eqref{full spectrum inside} we have a 
slightly different scenario, where oscillations are provided exclusively by the $t$-dependent exponential.
We can therefore easily split the function $\Omega^{II}$ into its real and imaginary part
\begin{eqnarray}
    \text{Re}[\Omega^{II}]=&&\cos(\omega t)e^{-r\sqrt{\mu^2-\omega^2}
-\frac{[\mu^2 M+e\omega Q-2M\omega^2]}{\sqrt{\mu^2-\omega^2}}\ln r} P^0_0(z,\gamma^2),\\
    \text{Im}[\Omega^{II}]=&&\sin(\omega t)e^{-r\sqrt{\mu^2-\omega^2}
-\frac{[\mu^2 M+e\omega Q-2M\omega^2]}{\sqrt{\mu^2-\omega^2}}\ln r} P^0_0(z,\gamma^2).
\end{eqnarray}
As $t$ grows, we obtain strongly oscillating functions in a finite 
interval of integration. However we consider, as already stated, the case in 
which $t$ is fixed and perform our analysis at spacelike infinity.
We start from the outside part of the spectrum, given by Eq. \eqref{full 
spectrum outside}, by considering some plots (cf. Fig. \ref{fig1} and 
Fig. \ref{fig2}) of the real part of $\Omega^I(r,z,\gamma^2)$ in function of $r$, fixed 
$z$ and $\omega$. As discussed in the captions, as $\omega$ increases, no matter the 
values of $z$ or of the other coefficients, the function exhibits stronger oscillations 
with increasing amplitude. This suggests that the overall contribution of the integral 
of Eq. \eqref{full spectrum outside} just vanishes, since we are integrating 
a strongly oscillating function. 

The case in which $\omega^2<\mu^2$, given by Eq. \eqref{full spectrum inside}, provides 
instead the opportunity to go further in analytical approximations.
For small values of $\gamma^2$, achieved for instance by considering a slowly 
rotating black hole ($a\ll1$), we can write explicitly 
that the spheroidal wave functions 
of the first and second kind take the form  
\begin{eqnarray}
    Ps^0_0(z,\gamma^2)\simeq&& 1+\frac{1}{18}(1-3z^2)\gamma^2
+\frac{(135z^4-30z^2-37)}{16200}\gamma^4 +O(\gamma^{6}),\\
    Qs^0_0(z,\gamma^2)\simeq&&\frac{1}{2}\ln\frac{(1+z)}{(1-z)}
+\frac{1}{36}\left(-12z+\ln\frac{(1+z)}{(1-z)}
+3z^2\ln\frac{(1-z)}{(1+z)}\right)\gamma^2+O(\gamma^4),
\end{eqnarray}
respectively. It is known from the theory that 
the full solution (which we rewrite as a 
reminder) $P^0_0(z,\gamma^2)=c_1  Ps^0_0(z,\gamma^2)+c_2Qs^0_0(z,\gamma^2)$ is 
bounded in the interval $z\in]-1,1[$, henceforth the above series expansion is 
convergent. This allows us to integrate it term by term, where every term is an integral 
of an exponent times a polynomial form in $\omega$. We stress that, since each term has 
its own $a^n$ coefficient, we can read them as progressive perturbations of the spherical case.
The exponent instead can be simplified by considering the case of a small massive field, 
i.e., $\mu\ll1$. Upon considering this scenario we are able to compute the integral 
term by term, reaching the desired level of accuracy.

\subsection{An application: resonant frequencies of quasibound states}

In Ref. \cite{VBM2022}, the computation of frequencies of quasibound states 
relies upon using the polynomial condition of the Heun function as a matching 
procedure for the two (asymptotic) radial solutions in their common overlap 
region. The asymptotic behaviour of the field at spacelike infinity is taken 
to be (with our notation)
\begin{equation}
    R_\infty(r)\sim \mathcal{N} \frac{e^ {-r \sqrt{-A_{0}}} \; r^\zeta}{r},
\end{equation}
which corresponds to a kind of large-$r$ approximation of our solution in Eq. 
\eqref{our full radial solution}. However, their analysis relies on the Heun 
polynomial condition, hence they just use the Heun parameters without considering 
the explicit form of the solutions at the event horizon and spacelike infinity.

\section{Massive fermionic field in Kerr-Newman space-time}
\label{A massive fermionic field in Kerr-Newman space-time}

The study of the Dirac equation in fixed curved space-times has a long and well 
established tradition. We would like to mention the seminal works of Chandrasekhar 
\cite{Chandra1976} and Page \cite{Page1976}, who separated such an 
equation for a massive field in the Kerr-Newman space-time. Since the procedure 
is more involved with respect to the Klein-Gordon equation, we give just a brief 
overview of the derivation of the purely radial equation, referring to the extensive 
literature on this topic for further details. 

\subsection{A brief introduction to the field equation}

In order to outline the derivation of our desired equation, we follow Ref. 
\cite{Kraniotis2019}, where it is obtained via the Newman-Penrose null tetrad 
formalism (a slightly different approach is considered, e.g., in 
Ref. \cite{Dariescu2021}, where the authors exploit the symmetries of 
the full Dirac equation). 
\newline
It is well known that the Kerr-Newman geometry with metric
\begin{equation}
\begin{split}
ds^2=&\left(\frac{\Delta  - a^2\sin^2\theta}{\rho^2}\right)dt^2
+\frac{2a(r^2+a^2-\Delta)\sin^2\theta}{\rho^2}dtd\phi \\& -\left[\frac{(r^2+a^2)^2
-\Delta a^2\sin^2\theta}{\rho^2}\right]\sin^2\theta d\phi^2
+\frac{\rho^2}{\Delta}dr^2+\rho^2d\theta^2,
\end{split}
\end{equation}
with $\rho^2$ and $\Delta$ given by Eqs \eqref{rhoquadro kerr-newman bl coordinates}, 
\eqref{Delta for the kerr newman metric}, respectively,
can be described in terms of a local 
Newman-Penrose null tetrad, where the tetrad coincides with the two principal 
null directions of the Weyl tensor (since the metric, as most black hole metrics, is 
of Petrov type D). Therefore the starting point is given by the 
generalised Kinnersley tetrad
\begin{eqnarray}
l^\mu=&&\left(\frac{(r^2+a^2)}{\Delta},1,0,\frac{a}{\Delta}\right),\\
n^\mu=&&\left(\frac{(r^2+a^2)}{2\rho^2},
-\frac{\Delta}{2\rho^2},0,\frac{a}{2\rho^2}\right),\\
m^\mu=&&\frac{1}{\sqrt{2}(r+ia\cos\theta)}\left(ia\sin\theta,0,1,
\frac{i}{\sin\theta}\right).
\end{eqnarray}
The evaluation of Ricci coefficients is then necessary in order to write the 
following 2-spinor form of the Dirac equation (cf. Eqs $(19)$-$(20)$ of Ref.
\cite{Kraniotis2019}, where several other details are given):
\begin{eqnarray}
(\nabla_{A{\dot B}}+ieA_{A{\dot B}})P^{A}
+i\frac{\mu}{\sqrt{2}}
{\bar Q}_{\dot B}&&=0,\\
(\nabla_{A{\dot B}}-ieA_{A{\dot B}})Q^{A}
+i\frac{\mu}{\sqrt{2 }}{\bar P}_{\dot B}&&=0.
\end{eqnarray}
Upon considering the factorization ansatz 
\begin{align}
P^{(0)} &=\frac{1}{\sqrt{2}(r-ia\cos\theta)} e^{- i \omega t + i m \phi}\,
S^{(-)}(\theta)\, R^{(-)}(r), 
\\
P^{(1)} &=\frac{1}{\sqrt{\Delta}} e^{- i \omega t + i m \phi}\,
S^{(+)}(\theta)\, R^{(+)}(r),
\end{align}

\begin{align}
\bar Q^{(\dot 0)} &= \frac{1}{\sqrt{2}(r+ia\cos\theta)}e^{- i \omega t + i m \phi}\,
S^{(+)}(\theta)\, R^{(-)}(r), 
\\
\bar Q^{(\dot 1)} &=\frac{1}{\sqrt{\Delta}} e^{- i \omega t + i m \phi}\,
S^{(-)}(\theta)\, R^{(+)}(r),
\end{align}
the Dirac equation reduces to a set of coupled differential equations, 
two for the radial parts 
\begin{equation}
\sqrt{\Delta}\Bigg[
\frac{d R^{(-)}(r)}{dr}
+
\left\{
\frac{i (m a - \omega \left(r^{2}+a^{2}\right))}{\Delta}
\right\}
R^{(-)}(r)
-
\frac{i\,Q e\, r}{\Delta}
R^{(-)}(r)
\Bigg]
=
(\lambda + i \mu r)\, R^{(+)}(r), 
\label{first combined radial equation massive dirac kernewman}
\end{equation}
\begin{equation}
\sqrt{\Delta}\,
\frac{d R^{(+)}(r)}{dr}
-
\frac{i\!\left(m a - \omega \left(r^{2}+a^{2}\right)\right)}{\sqrt{\Delta}}
R^{(+)}(r)
+
\frac{i\,Q e\, r}{\sqrt{\Delta}}
R^{(+)}(r)
=
(\lambda - i \mu r)\, R^{(-)}(r), 
\label{second combined radial equation massive dirac kernewman}
\end{equation}
(where $\Delta$ is defined in Eq. \eqref{Delta for the kerr newman 
metric}) and two for the angular parts
\begin{align}
\frac{d S^{(+)}(\theta)}{d\theta}
&+
\left[
\frac{m}{\sin\theta}
-
\omega a \sin\theta
+
\frac{\cot\theta}{2}
\right]
S^{(+)}(\theta)
=
\left(
-\lambda + \mu a \cos\theta
\right)
S^{(-)}(\theta),
\\
\frac{d S^{(-)}(\theta)}{d\theta}
&+
\left[
- \frac{m}{\sin \theta}
+ \omega a \sin\theta
+
\frac{\cot\theta}{2}
\right]
S^{(-)}(\theta)
=
\left(
\lambda + \mu a \cos\theta
\right)
S^{(+)}(\theta).
\end{align}
Note that Eq. \eqref{first combined radial equation massive dirac kernewman} 
can be obtained from Eq. \eqref{second combined radial equation massive dirac 
kernewman} by a complex conjugation. The analysis of the author 
of Ref. \cite{Kraniotis2019} leads to the result 
that the four functions $R^{(\pm)}(r), S^{(\pm)}(\theta)$ obey generalised Heun 
equations which give the solution near the singular points and 
the one at $r\rightarrow \infty$ for the radial part (which we will compare with 
our result below). Furthermore, the separation term $\lambda$ is extensively 
studied in his Sec. $9$. This is the level of detail we need in order to apply 
the Poincar\'e series expansion for the radial field given 
by the functions $R^{(\pm)}(r)$.

\subsection{Asymptotic expansion of the radial part}

Once that the radial equation has been obtained and its profile has been established
(in this case we are dealing with a generalised Heun equation), we can safely apply 
the formulae given in Sec. \ref{Poincare asymptotic series expansion}. As already 
stated, the radial part of the field consists of two components, 
\begin{equation}
R(r)=    \begin{pmatrix}
R^{(+)}(r) \\
R^{(-)}(r)
\end{pmatrix},
\end{equation}
and we begin by considering the $R^{(-)}(r)$ part. The second-order radial equation 
can be obtained from Eqs 
\eqref{first combined radial equation massive dirac kernewman} and 
\eqref{second combined radial equation massive dirac kernewman}:
\begin{equation}
    \left[\frac{d}{dr^2}+p^{(-)}(r)\frac{d}{dr}+q^{(-)}(r)\right]R^{(-)}(r)=0,
    \label{second order radial equation obeyed by R^{(-)}(r)}
\end{equation}
with
\begin{eqnarray}
    p^{(-)}(r)=&&\frac{1}{\Delta}\left(r-M-\frac{i\mu\Delta}{(\lambda+i\mu r)}\right),\\
    q^{(-)}(r)=&&\frac{1}{\Delta}\left[\frac{(K^2+iK(r-M))}{\Delta}
-2i\omega r-ieQ-\frac{\mu K}{(\lambda+i\mu r)}-\mu^2 r^2-\lambda^2\right],
\end{eqnarray}
where $K=(r^2+a^2)\omega -ma+erQ$.
Upon setting 
\begin{equation}
R^{(-)}(r)=\alpha^{(-)}(r)\beta^{(-)}(r),
\end{equation}
we obtain as a first step the expression for the first factor,
\begin{equation}
\alpha^{(-)}(r)=e^{-\frac{1}{4}\ln(2Mr-r^2-Q^2-a^2)}e^{\frac{1}{4}
\ln(\mu^2r^2+\lambda^2)}e^{\frac{i}{2}\arctan\left(\frac{\mu r}{\lambda}\right)}.
\end{equation}
Note that, since we are investigating the exterior of the black hole 
space-time, i.e., what we have called region III in Subsec. \ref{The ker 
newman black hole, subsection} defined by $r>r_+$ (with $r_+$ defined in 
Eq. \eqref{two horizons of the kerr newman metric}), the argument of the first 
logarithm is always a negative number. We can write it explicitly by considering 
the formula for the logarithm of a negative number and taking
the principal value, obtaining
\begin{equation}
    \ln(2Mr -r^2-Q^2-a^2)=\ln(r^2-2Mr+Q^2+a^2)+i(\pi+2\pi k) \quad \text{with }\quad k=0.
\end{equation}
Therefore we can safely write
\begin{equation}
\alpha^{(-)}(r)=\left(1-\frac{i}{\sqrt2}\right)\left(
\frac{\mu^2r^2+\lambda^2}{r^2-2Mr+Q^2+a^2}\right)^{\frac{1}{4}}
e^{\frac{i}{2}\arctan\left(\frac{\mu r}{\lambda}\right)},
\end{equation}
obtaining a term that on this occasion is rather more involved with respect to the scalar 
case given by Eq. \eqref{alpha part radial canonical solution}.  
The canonical potential is given by
\begin{align}
J^{(-)}(r)
={}&
\frac{(r-M)\,(2r-2M)}{2\Delta^{2}}
-
\frac{
\displaystyle
\frac{i\mu\Delta}{(\lambda+i\mu r)}\,(2r-2M)
}
{2\Delta^{2}}
\nonumber \\[1.0em]
&+
\frac{
[-(Q^{2}+a^{2})\mu^{2}
+ 2 i M \lambda \mu
+ \lambda^{2}]
}
{2\,(2Mr-Q^{2}-a^{2}-r^{2})\,(\lambda+i\mu r)^{2}}
\nonumber \\[1.0em]
&-
\frac{1}{4\Delta^{2}}
\left(
r-M
-
\frac{i\mu\Delta}{(\lambda+i\mu r)}
\right)^{2}
\nonumber \\[1.0em]
&+
\frac{1}{\Delta}
\Bigg[
\frac{
\left((a^{2}+r^{2})\omega - ma + Q e r\right)^{2}
+ i(r-M)\left((a^{2}+r^{2})\omega - ma + Q e r\right)
}
{\Delta}
\nonumber \\[0.6em]
&\qquad
- 2 i \omega r - i e Q
- \frac{
\mu\left((a^{2}+r^{2})\omega - ma + Q e r\right)
}
{(\lambda+i\mu r)}
\nonumber \\[0.6em]
&\qquad
- \mu^{2} r^{2}
- \lambda^{2}
\Bigg],
\end{align}
and we can expand it at $r\rightarrow \infty$, obtaining for the first 
coefficients $A_k^{(-)}(\omega,\mu,e,a,Q,M)$ (cf. Eq. \eqref{(6.2)})
\begin{align}
A_0^{(-)} &= \omega^2 - \mu^2, \\[6pt]
A_1^{(-)} &= 2 \omega e Q
            + 2 \left( 2 \omega^2 - \mu^2 \right), \\[8pt]
A_2^{(-)} &= \frac{1}{\mu} \Bigl[
   \left( -4 M^2 + Q^2 + a^2 \right) \mu^3
   + \mu \Bigl(
       (12 M^2 - 2 Q^2) \omega^2 \nonumber\\
&\qquad\qquad
       + (8 M Q e + i M - 2 m a)\,\omega
       + Q^2 e^2
       + i e Q
       - \lambda^2
     \Bigr)
   - \omega \lambda
\Bigr], \\[10pt]
A_3^{(-)} &= \frac{1}{\mu^2} \Bigg[
   4 \bigl( -2 M^3 + M (Q^2 + a^2) \bigr) \mu^4 \nonumber\\[4pt]
&\quad
   + \mu^2 \Bigl(
       32 M^3 \omega^2
       + 4 \omega (6 e Q + i) M^2 \nonumber\\
&\qquad\qquad
       + M \bigl(
           1
           + 4 (-3 Q^2 - a^2) \omega^2
           - 8 a m \omega
           + 4 Q^2 e^2
           + 3 i e Q
           - 2 \lambda^2
         \bigr)\nonumber \\
&\qquad\qquad
       + \bigl(
           -4 Q^3 e
           - 2 Q a^2 e
           - i Q^2
           + i a^2
         \bigr) \omega
       - 2 m (e Q + i) a
     \Bigr)\nonumber \\[4pt]
&\quad
   - (2 M \omega + e Q + i)\,\lambda \mu
   - i \lambda^2 \omega \Bigg]. 
\end{align}
We note that the $A_0$ coefficient is the same as the one for the massive 
scalar field, given by Eq. \eqref{A_0 massive ker newman}. By following Eqs 
\eqref{zeta general expression} and \eqref{B1 general expression}, we 
have that the power $\zeta^{(-)}$ of the $r$ term and the 
$B_1^{(-)}(\omega,\mu,e,a,Q,M)$ coefficient of the asymptotic expansion 
(cf. Eq. \eqref{(6.2)}) are given by
\begin{align}
\zeta^{(-)} &=
\frac{
  [2 \omega^2 M
  + \omega e Q
  - M \mu^2]
}{
  \sqrt{\mu^2 - \omega^2}
}, \nonumber\\[10pt]
B_1^{(-)} &=
\frac{1}{
  2 \mu (\mu - \omega)^2 (\mu + \omega)^2
}
\Bigg[
  \Bigl(
      (3 M^2 - Q^2 - a^2) \mu^5 \nonumber\\[2pt]
&\qquad
    + \mu^3 \Bigl(
        (-12 M^2 + 3 Q^2 + a^2) \omega^2
        + \bigl( (-6 e Q - i) M + 2 m a \bigr) \omega
        - Q^2 e^2
        - i e Q
        + \lambda^2
      \Bigr)\nonumber \\[2pt]
&\qquad
    + \lambda \mu^2 \omega \nonumber\\[2pt]
&\qquad
    + \mu \omega^2 \Bigl(
        (8 M^2 - 2 Q^2) \omega^2
        + \bigl( (4 e Q + i) M - 2 m a \bigr) \omega
        + i e Q
        - \lambda^2
      \Bigr)\nonumber \\[2pt]
&\qquad
    - \lambda \omega^3
  \Bigr)
  \sqrt{\mu^2 - \omega^2} \nonumber\\[6pt]
&\quad
  - \mu (\mu - \omega)(\mu + \omega)
    \bigl(
      M \mu^2
      - 2 M \omega^2 \Bigl) \Bigg].
\end{align}
The above results are all we need in order to construct the full radial 
asymptotic solution given by Eq. \eqref{full radial asymptotic solution} 
for the $R^{(-)}(r)$ part of the radial fermionic field. The $R^{(+)}(r)$ 
part is easily obtained by a complex conjugation (labelled with $^*$), since 
its equation is of the same form of Eq. \eqref{second order radial equation 
obeyed by R^{(-)}(r)} where
\begin{equation}
    p^{(+)}(r)=[p^{(-)}(r)]^*, \quad q^{(+)}(r)=[q^{(-)}(r)]^*,
\end{equation}
and therefore all the other quantities follow in a straightforward way.

\subsubsection{Addendum: massive and massless case}

The $A_s^{(-)}$ and $B_s^{(-)}$ coefficients, previously computed, diverge 
in the $\mu\rightarrow 0$ limit, and hence the massless case (e.g., the study 
of a massless neutrino) cannot be studied in such a way. However, this issue 
is not a drawback of our framework, but is instead largely recognized in the study 
of the dynamics of fields in curved backgrounds. One of the clearest examples 
is provided by DeWitt (cf. Ref. \cite{DeWitt1984} for a pedagogical introduction, 
Ref. \cite{DeWitt2003} for a global monograph). As a representative case study, 
we mention the quantization of linear massive bosons fields, when the Green function 
given in Eq. $(9.100)$ in Ref. \cite{DeWitt1984} contains a series of negative powers 
of the field mass. Such a behaviour holds also in other scenarios. Therefore, the 
massless case requires a separate analysis. For instance, one has to consider the 
$\mu\rightarrow 0$ limit directly in Eq. \eqref{second order radial equation obeyed by R^{(-)}(r)}, 
before computing the canonical reduction and then the relevant coefficients. 
A comparison can then be performed. In general, the study of the massless case relies 
on the resolvent of a hyperbolic differential operator on a manifold. The same 
argument holds for the other fields studied in this work.

\subsubsection{Analysis of the leading term}

The leading term of our solution for $R^{(-)}(r)$ is given by neglecting the 
interpolating series in Eq. \eqref{full radial asymptotic solution}. 
It reads explicitly as
\begin{equation}
R^{(-)}(r)\simeq \left(1-\frac{i}{\sqrt2}\right)\left(
\frac{\mu^2r^2+\lambda^2}{r^2-2Mr+Q^2+a^2}\right)^{\frac{1}{4}}
e^{\frac{i}{2}\arctan\left(\frac{\mu r}{\lambda}\right)}e^{-r\sqrt{\mu^2-\omega^2}} 
r^{\frac{[2 \omega^2 M+ \omega e Q- M \mu^2]}{ \sqrt{\mu^2 - \omega^2}} \nonumber}.
\label{leading order massive dirac field radial kerr newman}
\end{equation}
The qualitative behaviour is dictated by the piece $\sqrt{\mu^2-\omega^2}$: 
when $|\omega|>\mu$, the last two terms exhibit higher-in-frequency oscillations 
as $r$ grows, while for $|\omega|<\mu$ the damping exponential 
$e^{-r\sqrt{\mu^2-\omega^2}}$ controls their amplitude. In both cases, the prefactor 
reduces to $\sqrt{\mu}$ in the limit $r\rightarrow\infty$, highlighting the role 
of the field's mass in its overall trend. The $\omega=\mu$ case is singular, 
however, it does not represent a problem since it is just a zero-measure 
set in the integration over the full spectrum of the total solution. It is 
interesting to notice, in conclusion, that other quantities, such as the 
extremality of the black hole or the separation variable $\lambda$, do 
not play a relevant role at this stage.

\subsubsection{A plot of the solution}

\begin{figure}
    \centering
    \includegraphics[width=0.5\linewidth]{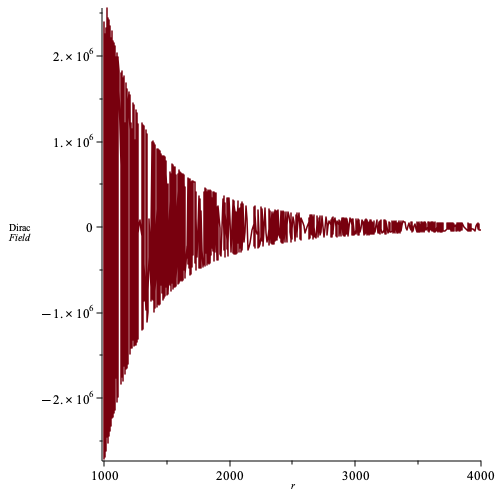}
\caption{The real part of the Dirac asymptotic solution (cf. Eq.\eqref{full 
radial asymptotic solution}) in the interval $r\in[10,40]r_+$, for the parameters 
given in Eqs \eqref{paramaters choice for plot massive dirac field}. While 
the frequency of oscillations remains stable, their amplitude decreases as 
$r$ grows. This behaviour is very similar to the one investigated for a massive 
scalar field in the same space-time.}
    \label{Fig3}
\end{figure}

By considering Eq. \eqref{full radial asymptotic solution} with $\mathcal{N}=1$ 
and the following values for the parameters (corresponding to 
a non-extremal Kerr-Newman black hole)
\begin{equation}
    M=100, \quad Q=5, \quad a=30, \quad e=10, \quad \mu=3, \quad m=10, 
\quad \lambda\simeq 1, \quad \omega=5,
    \label{paramaters choice for plot massive dirac field}
\end{equation}
we obtain the plot in Fig. \ref{Fig3} for the real part of the field (the one 
for the imaginary part is very similar), showing damped oscillations at spacelike 
infinity, i.e., $r\gg r_+$.  By choosing the range $|\omega|<\mu$ for the field 
frequency, we have instead that the field vanishes. Such results confirm our 
previous analysis of the leading order of the series. 

\subsubsection{Full spectrum analysis}

The leading order of our solution, given by Eq. \eqref{leading order massive 
dirac field radial kerr newman}, makes it possible for us to compute the total solution of 
the Dirac equation over the whole spectrum. Even if the angular part is more 
involved with respect to the scalar case (cf. Ref. \cite{Kraniotis2019}), the 
method remains the same: we can expand the angular part in powers 
of $a$, treating the axial symmetry as a deformation of the spherical one. We 
therefore obtain a series which can be integrated term by term (thanks to the 
convergence properties of spheroidal functions).

\section{Spin-weighted solutions of the Teukolsky equation}
\label{Spin-weighted solutions of Teukolsky equation}

In this section we provide the explicit computation of the coefficients of 
the asymptotic series solution (cf. Eq. \eqref{full radial asymptotic solution}) 
in the case of a massless field with arbitrary spin in Kerr space-time. 
This case plays a pivotal role in the study of black hole solutions and 
perturbations by virtue of its historical importance and its wide range of 
applicability, from applied mathematics to astrophysics. The first one who 
approached such a problem was Teukolsky, who developed the method  
and performed a great deal of computations and 
applications (cf., e.g., Refs 
\cite{Teukolsky1973I,Teukolsky1973II,PT1973}) of his results. 

\subsection{Main aspects of the Teukolsky equation}

As explained in Sec \ref{Stationary, axial-symmetric black holes}, the existence 
of the principal tensor makes it possible to solve field equations on Petrov type-D metrics 
by separation of variables, casting this apparently accidental feature into a more 
general and fundamental frame. However, these considerations are only understood 
recently. Teukolsky showed that, in a Kerr space-time (admitting the possibility 
to extend this analysis also to other black hole metrics), one can obtain separable 
equations for the radiative parts of electromagnetic and gravitational perturbations. 
Moreover, they can be unified into a single 
``master" equation (known as Teukolsky 
master equation) with a spin-weight parameter $s$, describing the type of field 
under study (scalar, electromagnetic, gravitational). Here we report just the main 
steps, while the full analysis can be found in the references cited earlier. We 
consider the Kerr metric given by Eq. \eqref{kerr metric in bl coordinates}, studying 
perturbations in the Newman-Penrose formalism (which arises from the introduction of 
spinor calculus in General Relativity). After some manipulations, it is possible to 
write a master equation for a massless spin-weighted field $\psi^{(s)}$, which might 
be a scalar ($s=0$), neutrino ($s=\pm1/2$) or electromagnetic ($s=\pm1$) test field 
or, furthermore, a gravitational ($s=\pm2$) perturbation. Such an equation reads
\begin{equation}
\begin{split}
    &\left[\frac{(r^2+a^2)^2}{\Delta}-a^2\sin^2\theta\right]
\frac{\partial^2 \psi^{(s)}}{\partial t^2}+\frac{4Mar}{\Delta}
\frac{\partial^2 \psi^{(s)}}{\partial t \partial \phi}+\left[
\frac{a^2}{\Delta}-\frac{1}{\sin^2\theta}\right]
\frac{\partial^2 \psi^{(s)}}{\partial \phi^2}\\&-\Delta^{-s}
\frac{\partial}{\partial r}\left(\Delta^{s+1}
\frac{\partial \psi^{(s)}}{\partial r}\right)-\frac{1}{\sin\theta}
\frac{\partial}{\partial \theta}\left(\sin\theta 
\frac{\partial \psi^{(s)}}{\partial \theta}\right) -2s\left[
\frac{a(r-M)}{\Delta}+i\frac{\cos\theta}{\sin^2\theta}\right] 
\frac{\partial\psi}{\partial\phi}\\&-2s\left[\frac{M(r^2-a^2)}{\Delta}-r
-ia\cos\theta\right]\frac{\partial \psi}{\partial t}+(s^2\cot^2\theta-s)\psi=0,
\label{teukolsky initial equation all coordinates}
\end{split}
\end{equation}
with (rewritten for convenience) $\Delta=r^2-2Mr+a^2$ and vanishing source term. The ansatz
\begin{equation}
    \psi^{(s)}_{lm}(t,r,\theta.\phi)=e^{-i\omega t}e^{im\phi}S^{(s)}_{lm}(\theta)R^{(s)}(r)
    \label{separation variables teukolsky}
\end{equation}
turns the above master equation into a pair of decoupled equations. 
We postpone the description of the radial equation to the next section, while 
the angular one (cf. Ref. \cite{Teukolsky1973II} and references 
therein) is briefly described below.

\subsubsection{The angular equation}

Thanks to the factorization ansatz given in Eq. \eqref{separation variables teukolsky}, 
we obtain from Eq. \eqref{teukolsky initial equation all coordinates} the angular 
one, which can be written as an eigenvalue problem 
\begin{equation}
    (\mathbb{S}_1+\mathbb{S}_2)S^{(s)}(\theta)=-\lambda^s_{lm}(a\omega) S^{(s)}_{lm}(\theta)
    \label{angular equation Teukolsky}
\end{equation}
for the two differential operators
\begin{equation}
\begin{split}
\mathbb{S}_1=&\frac{1}{\sin\theta}\frac{d}{d\theta}\left(\sin\theta\frac{d}{d\theta}\right)
-\left(\frac{m^2+s^2+2ms\cos\theta}{\sin^2\theta}\right),\\
\mathbb{S}_2=&a^2\omega^2\cos^2\theta-2a\omega s\cos\theta.
\end{split}
\end{equation}
The eigenvalue problem for the first operator (independent of $a$)
\begin{equation}
    \mathbb{S}_1S^{(s)}_{lm}(\theta)=-\lambda^s_{lm}(a\omega) S^{(s)}_{lm}(\theta)
\end{equation}
is solved by spin-weighted spherical harmonics ${}_sY_{lm}(\theta,\phi)$, a spin 
generalization of the standard spherical harmonics. They can be written explicitly 
in terms of Jacobi polynomials 
\begin{equation}
   {}_sY_{lm}(\theta,\phi)= N_{lms} e^{im\phi}(1-x)^{(m+s)/2}(1+x)^{(m-s)/2}P_{l-m}^{(m+s,m-s)}(x),
\end{equation}
where $x=\cos\theta$ (frequently used in the mathematical literature) and 
$P_n^{(m+s,m-s)}$ are Jacobi polynomials ($n$ labels the degree of the polynomial, 
understood as $P_n^{(m+s,m-s)}=\sum_{i=0}^n P_i^{(m+s,m-s)}$).
The second angular operator $ \mathbb{S}_2$ encodes instead the departure from 
the spherical to the spheroidal case, by virtue of axial symmetry of space-time. 
It is usually treated with perturbative or numerical methods, while analytical 
ones refer to particular cases (cf., e.g., Ref. \cite{Chen:2022rrg}). Here, we 
are interested in the $a\omega=0, s\neq0, \omega\in\mathbb{R}$ case of Eq. 
\eqref{angular equation Teukolsky}, since it makes it possible for us to perform a full
spectrum analysis of the field at spacelike infinity for different values of its spin. 
In such a case, eigenvalues and eigenfunctions are simply
\begin{equation}
    \begin{split}
        \lambda^s_{lm}(0)=&l(l+1)-s(s+1),\\
        S^{(s)}_{lm}(x)=&N_{lms}(1-x)^{(m+s)/2}(1+x)^{(m-s)/2}P_{l-m}^{(m+s,m-s)}(x),
    \end{split}
\end{equation}
We recall that the smallest value $l$ is $l=\text{max}(|s|,|m|)$, while $m=0,\pm1,\pm2 ,
\dots$. In the literature they are often referred to as angular quantum number, for their 
property of being quantized. However, we stress that this nomenclature might be 
misleading since we are in a fully classical framework.

\subsection{Poincaré series for a spin-weighted massless field}

The second-order radial equation for $R^{(s)}(r)$, as written in Eq. 
\eqref{second order radial equation in the right form for poincare}, 
exhibits the following coefficients:
\begin{eqnarray}
    p^{(s)}(r)=&&\frac{2(s+1)(r-M)}{\Delta},\\
q^{(s)}(r)=&&\frac{1}{\Delta}\left[\frac{(K^2-2isK(r-M))}{\Delta}
+4ir\omega s-\lambda^s_{lm}(a\omega)\right],
\end{eqnarray}
where $K=(r^2+a^2)\omega$. In order to improve the readability of the 
following equations in this subsection, we label the separation variable as 
\begin{equation}
    \lambda^s_{lm}(a\omega)\equiv \lambda.
\end{equation}
The reduction to the canonical Laguerre-Forsyth form, based on the factorization
\begin{equation}
    R^{(s)}(r)=\alpha^{(s)}(r)\beta^{(s)}(r)
\end{equation}
and described in Sec. \ref{Poincare asymptotic series expansion}, yields for 
the $\alpha^{(s)}(r)$ piece (cf. Eq. \eqref{alpha part radial canonical 
solution}) the solution
\begin{equation}
    \alpha^{(s)}(r)=(2Mr-a^2-r^2)^{-\frac{1}{2}(s+1)},
\end{equation}
while the canonical potential (cf. Eq. \eqref{canonical series 
expansion potential}) reads as
\begin{equation}
J^{(s)}(r) =
\frac{1}{\left(r^{2}-2Mr+a^{2}\right)^{2}}
\begin{aligned}
\Bigg\{
&\;\omega\, r^{4}
+ 2 i \omega s\, r^{3}
\\[0.4em]
&+
\Big[
- s^{2}
- (6 i M \omega + 1)s
+ 2 a^{2} \omega
- \lambda^{2}
\Big] r^{2}
\\[0.4em]
&+
2 r
\Big[
M s^{2}
+ s \left(i a^{2} \omega + M\right)
+ M \lambda^{2}
\Big]
\\[0.4em]
&-
M^{2} s^{2}
+ a^{2}\left(-1 + 2 i M \omega\right)s
+ M^{2}
+ a^{2}\left(a^{2} \omega - ^{2} - 1\right)
\Bigg\}.
\end{aligned}
\end{equation}
The expansion of $J^{(s)}(r)$ near $r\rightarrow\infty$, i.e.
\begin{equation}
    J^{(s)}=\sum_{k=0}^{\infty}\frac{A_k^{(s)}}{r^k},
\end{equation}
yields the following expansion coefficients: 
\begin{eqnarray}
    A_0^{(s)}=&&\omega^2,\\
    A_1^{(s)}=&&4M\omega^2+2is\omega,\\
    A_2^{(s)}=&&12M^2\omega^2-\lambda^2-s^2+s(2iM\omega-1),\\
    A_3^{(s)}=&&32M^3\omega^2-2M(2a^2\omega^2+\lambda^2+s^2+s)-2ia^2\omega s,\\
A_4^{(s)} =&& 80 M^{4} \omega^{2} - 8 i M^{3} \omega\, s
+ \left(-24 a^{2} \omega^{2}- 4 \lambda^{2}- 5 s^{2}- 4 s+ 1 \right) 
M^{2}\nonumber \\&&- 2 i M a^{2} \omega\, s+ 2 a^{2}\left(s^{2}
+ \frac{1}{2}\lambda^{2}+ \frac{1}{2}s- \frac{1}{2}\right).\\\nonumber
\end{eqnarray}

From Eq. \eqref{zeta general expression}, the explicit form of the exponent of the 
$r$ term in the full radial solution \eqref{full radial asymptotic solution} is given by
\begin{equation}
    \zeta^{(s)}=\text{sign}(\omega)(2M\omega+is),
\end{equation}
where with sign(\dots) we denote the sign function. The $B^{(s)}$ coefficients 
are easily obtained, and for the first two we find, for instance,
\begin{align}
B_1^{(s)} ={}&
\frac{
\Bigl[\Bigl(
8 M^{2}\omega^{2}
-2 i M \omega s
-\lambda^{2}
-s
\Bigr)\sqrt{-\omega^{2}}
- i \omega s
-2 M \omega^{2}\Bigl]
}{2 \omega^{2}},
\\[2ex]
B_2^{(s)} ={}&
-\frac{1}{2\omega^{3}}
\Biggl\{
\sqrt{-\omega^{2}}
\Bigl[
2 M a^{2} \omega^{3}
+ i (4 M^{2}+a^{2}) s \omega^{2}
+ 3 M \Bigl(s^{2}-\tfrac{1}{3}\lambda^{2}-\tfrac{1}{3}s
+\tfrac{1}{3}\Bigr)\omega \nonumber \\[1ex]
&\hspace{2em}
- i s \Bigl(\lambda^{2}+s-\tfrac{1}{2}\Bigr)
\Bigr]
+\, 8 \omega
\Bigl[
2 M^{4} \omega^{4}
- i M^{3} s \omega^{3}
-\tfrac{M^{2}}{8}\bigl(4\lambda^{2}+s^{2}+4s-1\bigr)\omega^{2}\nonumber\\[0.5ex]
&\hspace{6em}
+\tfrac{i M}{8}(\lambda^{2}+s-4)s\omega
+\tfrac{\lambda^{4}}{32}
+\tfrac{\lambda^{2}s}{16}
+\tfrac{s^{2}}{8}
-\tfrac{\lambda^{2}}{16}
-\tfrac{s}{16}
\Bigr]
\Biggr\}. 
\end{align}
Notice how the vanishing value of the mass yields easier expression for the series expansion. 
The spin of the field enters just as a parameter in the study of the real part of the field.

\subsection{Full spectrum analysis}

In order to investigate the full spectrum of a spin-weighted test field, we consider 
the case (as previously mentioned) $a\omega=0\implies a=0$, i.e., Schwarzschild 
space-time. Spherical symmetry simplifies the problem by reducing the separation 
variable to simply a constant with respect to the frequency of the field (and hence
integration is straightforward). 
Moreover, we consider the leading radial term and $m=0$, 
which implies that $l=s$: it makes it possible to highlight the role of the spin of the field 
and to achieve mathematical control on the solution. By emphasizing the $\omega$-dependence 
in Eq. \eqref{separation variables teukolsky}, it reduces to
\begin{equation}
\psi^{(s)}_{s0}(\omega, t,r,x)=\mathbf{N}\left(\frac{1-x}{1+x}\right)^\frac{s}{2} 
P_s^{(s,-s)}(x)\frac{r^{is}}{[r(2M-r)]^{\frac{s+1}{2}}}e^{\varphi(\omega,t,r)},
\end{equation}
where $\mathbf{N}=\mathcal{N}N_{s0s}$ assembles
the normalization of the radial and angular parts, and the function
\begin{equation}
    \varphi(\omega,t,r)=-|\omega|(r-2M\ln r) -i\omega t
\end{equation}
encodes the frequency dependence. Notice that $\varphi(\omega,t,r)$, at spacelike 
infinity, leads to damped oscillations since the logarithmic divergence is suppressed 
by the $r$ term. Such a behaviour is confirmed by a direct computation, since the 
$\omega$ integration is straightforward, giving
\begin{equation}
    \text{Re}\left(\int e^{\varphi(\omega,t,r)}d\omega
\right)=\frac{[t\sin(\omega t)
+\cos(\omega t)(2M\ln r-r)]}{[(4M\ln r)(M\ln r-r)+r^2+t^{2}]}e^{\omega(2M\ln r-r)},
\end{equation}
for the real part, while the imaginary one reads
\begin{equation}
      \text{Im}\left(\int e^{\varphi(\omega,t,r)}d\omega\right)=i\frac{[t\cos(\omega t)
-\sin(\omega t)(2M\ln r-r)]}{[(4M\ln r)(M\ln r-r)+r^2+t^{2}]}e^{\omega(2M\ln r-r)}.
\end{equation}
We have therefore that the field does not exhibit divergences at spacelike infinity. 
Such a procedure could be extended also to other angular values, since the behaviour 
in frequencies is completely independent of them.

\section{An application to quasinormal modes}
\label{An application to quasinormal modes}

One of the most compelling research areas in black hole physics studies 
characteristic frequencies of oscillation of perturbed black holes 
(i.e., the dynamics of fields on black hole metrics). In analogy with the 
study of atoms and molecules, this area is called black hole spectroscopy (see 
Sec. 4.3 in Ref. \cite{Compere2018} for an introduction). In this framework, 
a pivotal role is covered by the so-called quasinormal modes: free 
modes of oscillation of a black hole obtained by solving the radial equation 
for a field in a black hole space-time under specific boundary conditions.  
There exists a wide and rapidly evolving literature 
on quasinormal modes. We rely on 
the review in Ref. \cite{Berti2009} (and references therein) for an introduction. 
In order to define them, we exploit the case of a massive scalar field on 
Schwarzschild space-time, where for convenience we write the metric as
\begin{equation}
g=-f(r) dt^{2}+\frac{dr^{2}}{f(r)}+r^{2}(d\theta^{2}+\sin^{2}\theta d\phi^{2}),
\label{metric schwarzschild with generic f(r) function}
\end{equation}
with $f(r)=1-2M/r$. By solving the Klein-Gordon equation via separation of 
variables \cite{ER}, we obtain for its radial part $R_{l\omega \mu}(r)$ the equation
\begin{equation}
\left[\frac{d^{2}}{dr^{2}}+\frac{2(r-M)}{r^{2}f(r)}\frac{d}{dr}
+\frac{1}{f(r)}\left(\frac{\omega^{2}}{f(r)}-\frac{l(l+1)}{r^{2}}
-\mu^{2}\right)\right]R_{l \omega \mu}(r)=0,
\label{radial equation in schwarzschild space-time massive scalar field}
\end{equation}
where $\mu$ labels the mass of the field. This equation 
can be cast in the standard form preferred in the literature by exploiting the  
tortoise coordinate $r_*$, defined as 
\begin{equation}
    r_*=\int dr f^{-1}(r)=r+2M\ln\left|\frac{r}{2M}-1\right|,
\end{equation}
obtaining
\begin{equation}
  \left[\frac{d^2}{dr_*^2}+\omega^2-V(r_*)\right]R_{l \omega \mu}(r_*)=0,
  \label{radial equation starting for quasi normal modes}
\end{equation}
where 
\begin{equation}
    V(r_*)=f(r_*)\left[\frac{l(l+1)}{r_*^2}+\frac{f'(r_*)}{r_*}\right].
\end{equation}
We remark that, although we show this simple case, the form of  Eq. 
\eqref{radial equation starting for quasi normal modes} holds for several 
spin-weighted fields in various space-times, changing only the potential 
term $V(r_*)$. Quasinormal modes are defined as waves that cannot come from 
infinity nor can arise from the black hole. One could imagine them as waves 
oscillating around the event horizon of black holes, slowly sneaking 
at infinity or into the black hole, with characteristic frequencies which 
we have to compute. Mathematically, we have to impose boundary conditions
\begin{eqnarray}
r_{*} \rightarrow -\infty \Longrightarrow
R_{l \omega \mu}(r_*)\sim&& e^{-i\omega r_*}, \label{Boundary condition for QNM at the horizon}\\
r_{*} \rightarrow + \infty \Longrightarrow 
R_{l \omega \mu}(r_*)\sim&& e^{i\omega r_*},\label{boundary condition for QNM at infinity}
\end{eqnarray}
at the horizon $(r_*\rightarrow -\infty)$ and at spacelike infinity 
$(r_*\rightarrow \infty)$, respectively. Quasinormal modes do not form a complete set 
of wave functions and are often regarded as quasistationary states, since 
waves can escape either to infinity or into the black hole. The system 
must be considered as a dissipative one. By imposing the boundary conditions 
given in Eqs \eqref{Boundary condition for QNM at the horizon},
\eqref{boundary condition for QNM at infinity} on the solution of Eq. 
\eqref{radial equation starting for quasi normal modes}, we obtain a countable 
infinity of quasinormal modes, labelled by the corresponding eigenfrequencies 
$\omega_{\text{QNM}}$, which have a real and an imaginary part. Computing quasinormal 
modes reduces to the determination of such a spectrum. 
There are several techniques to obtain it, from numerical analyses to 
more analytical ones, e.g., one could use Green functions theory (cf. Ref. 
\cite{Berti2006}) or the WKB approximation (as in Ref. \cite{Decanini2009}).

Our framework, based on asymptotic series expansion, 
can be in principle adopted in several contexts. In order to give an 
introduction to these possibilities, we investigate an application 
that exploits the continuous fractions method.

\subsection{The continuous fractions method}

One of the most reliable methods is the one proposed by Leaver \cite{Leaver1985}, 
based on continuos fractions. We briefly describe his derivation regarding a 
massless spin-weighted field in Schwarzschild space-time. Upon considering a 
rescaled radial coordinate $\tilde r=\frac{r}{2M}$, so that the event horizon 
is located at $\tilde r=1$ and the spin parameter $\varepsilon=s^2-1$, where 
$s$ labels the spin, Eq. \eqref{radial equation in schwarzschild space-time 
massive scalar field} reads
\begin{equation}
    \left[\tilde r(\tilde r-1)\frac{d^2}{d\tilde r^2}+\frac{d}{d\tilde r} 
-\left(\frac{\rho^2\tilde r^3}{\tilde r-1}+l(l+1)-\frac{\varepsilon}{\tilde r}
\right)\right]R_{l \rho\varepsilon}(\tilde r)=0,
\end{equation}
where $\rho=-i\omega$. Eqs \eqref{Boundary condition for QNM at the horizon}, 
\eqref{boundary condition for QNM at infinity} read
\begin{align}
    R_{l \rho\varepsilon}(\tilde r)\sim& (\tilde r-1)^\rho,
\label{boundary condition QNM horizon Leaver}\\
    R_{l \rho\varepsilon}(\tilde r)\sim &\tilde r^{-\rho} e^{-\rho \tilde r}.
\label{boundary condition QNM infinity Leaver}\\\nonumber
\end{align}
A solution which satisfies the horizon boundary condition is
\begin{equation}
    R_{r\varrho \varepsilon}(\tilde r)=(\tilde r-1)^\rho \tilde r^{-2\rho} 
e^{-\rho (\tilde r-1)}\sum_{n=0}^\infty a_n\left(\frac{\tilde r-1}{\tilde r}\right)^n.
    \label{Series Leaver QNM}
\end{equation}
By inserting this series into the radial equation, we find that the $a_n$ 
coefficients are defined by the three-terms recurrence relation (with initial condition $a_0=1$)
\begin{align}
    \alpha_0 a_1+\beta_0 \alpha_0=&0, \label{Leaver recurrence relation n 0}\\
    \alpha_n a_{n+1}+\beta_na_n+\gamma_n a_{n-1}=&0, \quad n=1,2\dots 
\label{Leaver recurrence relation n greater than 0}
\end{align}
where the coefficients are simply given by
\begin{align}
    \alpha_n=&n^2+2(\rho+1)n +2\rho+1,\\
    \beta_n=&-[2n^2+2(4\rho+1)n+8\rho^2+4\rho+l(l+1)-\varepsilon],\\
    \gamma_n=&n^2+4\rho n +4\rho^2-\varepsilon-1.
\end{align}

The boundary condition at spatial infinity is satisfied for those values of 
$\omega=\omega_n$ for which the series in Eq. \eqref{Series Leaver QNM} is 
convergent. The large $n$ behaviour of the expansion coefficients indicates that 
the uniform convergence of the series is obtained only if, for $n\rightarrow \infty$,
\begin{equation}
\frac{a_{n+1}}{a_n}\rightarrow1-\frac{\sqrt{2\rho}}{\sqrt{n}}+\frac{\left(2\rho-\frac{3}{4}\right)}{n}+\dots .
\end{equation}
Such a condition is obtained only for eigenvalues of $\rho$ corresponding to 
quasimormal mode frequencies. If it holds, the ratio among two successive $a_n$ coefficients 
which satisfy the recurrence relations in Eqs \eqref{Leaver recurrence relation n 0}, 
\eqref{Leaver recurrence relation n greater than 0} is given by the continued fraction
\begin{equation}
\frac{a_{n+1}}{a_n}=\frac{-\gamma_{n+1}}{\beta_{n+1}-\frac{\alpha_{n+1} \gamma_{n+2}}{\beta_{n+2}
-\frac{\alpha_{n+2}\gamma_{n+3}}{\beta_{n+3}-\dots}}},
\end{equation}
which is usually written formally as
\begin{equation}
    \frac{a_{n+1}}{a_n}=\frac{-\gamma_{n+1}}{\beta_{n+1-}}\frac{\alpha_{n+1}\gamma_{n+2}}{\beta_{n+2}-}\dots.
\end{equation}
In order to obtain a characteristic equation for quasinormal mode frequencies we evaluate the previous 
equation at $n=0$ and we consider Eq. \eqref{Leaver recurrence relation n 0} as a $n=0$ 
boundary condition on the $\frac{a_1}{a_0}$ ratio. We obtain therefore two conditions that must be satisfied:
\begin{align}
    \frac{a_1}{a_0}=&-\frac{\beta_0}{\alpha_0},\\
    \frac{a_1}{a_0}=&\frac{-\gamma_{1}}{\beta_{1-}}\frac{\alpha_{1}\gamma_{2}}{\beta_{2}-}\dots\
\end{align}
By equating them, we arrive at an implicit characteristic equation for quasinormal frequencies,
\begin{equation}
    0=\beta_0-\frac{\alpha_0\gamma_1}{\beta_1-}\frac{\alpha_1\gamma_2}
{\beta_2-}\frac{\alpha_2\gamma_3}{\beta_3-}\dots,
    \label{continous fraction Leaver for quasinormal modes}
\end{equation}
Thanks to this 
algorithm, several values of $\omega_{\text{QNM}}$ for different $n,l,s$ 
are computed, in agreement with other methods.

\subsection{Our analysis of quasinormal modes}

In order to apply our mathematical scheme to this kind of computation, we 
recall first our results for a massive scalar field in Schwarzschild space-time 
derived in our previous paper \cite{ER}.
Starting from Eq. \eqref{radial equation in schwarzschild space-time massive 
scalar field}, the canonical reduction leads to the reduced equation 
\begin{equation}
\left[\frac{d^{2}}{dr^{2}}+J_{l\omega \mu}(r)\right]\beta_{l\omega \mu}(r)=0,
\label{canoncial equation massive field schwarzschild}
\end{equation}
where the canonical potential reads
\begin{align}
J_{l\omega \mu}(r)=& \frac{1}{4r^{2}}+\frac{1}{4(r-2M)^{2}}
-\frac{[\frac{1}{2}+l(l+1)]}{r(r-2M)}
\nonumber \\
&+\frac{\omega^{2}r^{2}}{(r-2M)^{2}}
-\frac{\mu^{2}r}{(r-2M)}.
\end{align}
Quasinormal mode boundary conditions are given by (cf. Eqs (27), (49) of Ref. \cite{ER})
\begin{align}
r \rightarrow 2M \Longrightarrow
\beta_{l\omega \mu}(r)\sim&(r-2M)^{\frac{1}{2}-2iM\omega},\\
r \rightarrow \infty \Longrightarrow 
\beta_{l\omega \mu}(r)\sim&e^{r\sqrt{\mu^2-\omega^2}}
r^\frac{M(2\omega^2-\mu^2)}{\sqrt{\mu^2-\omega^2}},
\end{align}
instead of Eqs \eqref{boundary condition QNM horizon Leaver}, 
\eqref{boundary condition QNM infinity Leaver}. In the above expressions, 
we can set $\mu=0$ in order to study the massless case, finding therefore
the limiting behaviours
\begin{align}
r \rightarrow 2M \Longrightarrow
\beta_{l\omega 0}(r)\sim&(r-2M)^{\frac{1}{2}-2iM\omega},\\
r \rightarrow \infty \Longrightarrow
\beta_{l\omega 0}(r)\sim&e^{ir\omega}r^{2iM\omega}.
\end{align}
We obtain very similar expressions with respect to Leaver's ones, that we 
can further simplify through the rescaling
\begin{equation}
r = 2M \tilde r \quad \Omega=-2i\omega M,
\end{equation}
obtaining the expressions
\begin{align}
{\tilde r} \rightarrow 1 \Longrightarrow
\beta_{l\Omega 0}(\tilde r)\sim&(\tilde r-1)^{\frac{1}{2}+\Omega},\\
{\tilde r} \rightarrow \infty  \Longrightarrow 
\beta_{l\Omega 0}(\tilde r)\sim&e^{-\Omega \tilde r}\tilde r^{-\Omega},
\end{align}
which are formally very similar to Leaver's ones.
Since we already have a reduced radial equation 
in canonical form (which provides 
the foundation of Poincar\'e series expansion), in order to display
the relation with Leaver's radial function $\psi_l(\tilde r)$ (that we 
have rewritten as $R_{l \rho\varepsilon}(\tilde r)$ in order to adapt it 
to our notation), it is sufficient to notice that 
\begin{equation}
R_{l \rho\varepsilon}(\tilde r)=\sqrt{\frac{\tilde r}{(\tilde r-1)}}\beta_{l\omega 0}(\tilde r).
\label{relation our and leaver radial}
\end{equation}
We see therefore that our method easily reproduces the results obtained by Leaver and 
his continuous fractions, since both of them start with two local solutions of the radial 
equation analytically continued through an interpolating series. Even though the 
global convergence of the Poincar\'e asymptotic series is not formally proved, a truncation 
is needed in explicit computations, assuring a regular behaviour at least at leading orders. 
So far, the map in Eq. \eqref{relation our and leaver radial} establishes only a mathematical 
connection between the two frameworks, proving that further investigations in this research 
path are possible. However, detailed formal and numerical computations deserve a dedicated paper. 

\subsection{A brief discussion on more recent models}

In the previous sections, we analysed several classical black hole space-times in 
order to study the asymptotic behaviour of fields, where a physical result is 
provided by the computation of quasinormal modes. 
However, recent approaches try to generalise 
the space-time model in order to include semi-classical or quantum corrections induced, 
e.g., by the existence of a fundamental Planck length scale $\ell$. As discussed in 
Ref. \cite{Sucu:2026gro}, it is possible to study the fermionic dynamics and 
thermodynamics for a static and spherically symmetric black hole in the context of 
$f(R)$ theories of gravity. Such a black hole is described by a Schwarzschild-like 
metric (cf. Eq. \eqref{metric schwarzschild with generic f(r) function}) where
\begin{equation}
f(r)=1-\frac{2M}{r}+\sum_{n=1}^3a_{2n}\left(\frac{\ell}{r}\right)^{2n}.
\end{equation}
The correction orders are governed by the dimensionless parameters $a_{2n}$. They 
become relevant in near-horizon analyses (where quantum corrections are supposed to be 
relevant), while the asymptotic structure of the space-time resembles the classical one. 
As discussed in the paper, it is possible to study the propagation of a Dirac field in 
such a space-time. Moreover, the computation of 
quasinormal modes through the WKB approximation 
leads to a spectrum whit an overall structure qualitatively similar to the classical case. 
The effect of quantum corrections can be summarised as a small quantitative shift in 
both the real and imaginary part of $\omega_{QNM}$. Another interesting case study is 
provided by charged black holes in $f(R)$ gravity with non-linear electrodynamics (cf. 
Refs. \cite{Sucu:2026rsr, Sucu:2026hfc}), where the field Lagrangian has the form
\begin{equation}
\mathcal{L}(F)=a_0+F+aF^p, \quad F\equiv\frac{1}{4}F_{ab}F^{ab}
\label{nonlinear electrodynamics}
\end{equation}
and the space-time has still the form of Eq. 
\eqref{metric schwarzschild with generic f(r) function} with
\begin{equation}
f(r)=1-\frac{2M}{r}+\frac{Q^2}{r^2}+[2b(p-1)-1]\frac{2^{1-p}}{(3-4p)}aQ^{2p}r^{2-4p}.
\end{equation}
In such a formula, $M$ and $Q$ are the black hole mass and charge, while $b$ is a 
constant. The  $a$, $p$ terms denote the non-linear electrodynamics coupling and the 
power-law index, respectively, as defined in Eq. \eqref{nonlinear electrodynamics}. 
Non-linear parameters induce a small shift in the 
quasinormal mode spectrum of a Dirac field 
in such a space-time, leaving unchanged the overall behaviour. Further interesting 
cases can be found, e.g., in Refs. \cite{Sucu:2026andp, Gecim2020, Gecim:2013mfa}. 
In addition to classical gravity, it is therefore possible to apply the Poincar\'e 
series also in semiclassical and/or quantum models of space-times, provided that the 
dynamics of fields is still defined (as in the mentioned cases). Such considerations 
suggest another possible application of our mathematical framework to non-trivial scenarios.

\section{Conclusions}

In this paper we have applied the Poincar\'e series expansion, investigated 
for the first time in our previous paper \cite{ER}, throughout the most 
important equations in black hole physics: Klein-Gordon, Dirac and Teukolsky 
equations for massive and massless fields in Kerr-Newman and Kerr space-times. 
By offering a broad overview of such space-times, we highlight how the 
method relies on the possibility to separate the equations of motion
into a radial and an angular part: this is achieved by virtue of 
the existence of a principal tensor, a deep geometrical object strictly related 
to the concept of hidden symmetries emerging in Petrov D metrics. We have 
devoted a large part of the paper to the description of these features, since they provide 
the underlying mathematical structure to exploit the series expansion. We then 
perform explicit computations of the coefficients of the series for a massive 
scalar and fermionic field in Kerr-Newman space-time: this allows us to deeply 
investigate the structure of the radial equation and perform an analysis not only 
of the fixed-frequency Fourier mode, but also of the whole spectrum. We then 
investigate spin-weighted massless solutions of the Teukolsky equation, by arriving 
also in this case to a full spectrum analysis. 
This topic deserves further remarks. In the literature, what is often studied is 
the single Fourier mode of the spectral decomposition of a field, leaving the whole wave 
packet as a formal integral. However, the latter is the true physical quantity which 
should be considered. A formal mathematical study of the whole spectrum, based on 
functional analysis, is provided in Ref. \cite{Huh} and largely discussed in Sec. VII 
of our paper \cite{ER} for a massive scalar field in Schwarzschild space-time. We 
briefly mention some of our considerations for this case study. In computing the whole 
Fourier integral, we face a case where a field not only does not vanish at spacelike 
infinity, but furthermore it diverges exponentially. Such a property is 
already recognized in Ref. \cite{Huh}, since part (II) of Proposition 2.1 therein shows 
that the norm of a massive scalar field
obeying the wave equation in Schwarzschild space-time is majorized by the product of
two functions, one of which is indeed divergent in a suitable limit. This agreement merits
explicit mention because the work in Ref. \cite{Huh} has exploited advanced methods from the
modern theory of hyperbolic equations, whereas we have used the standard techniques of
classical mathematical physics, supported by numerical computations. To the best of our 
knowledge, this is one of the few cases when the study of the whole spectrum is attempted, 
providing a solid mathematical background for our results.
We conclude with a first attempt to 
study quasinormal modes, preparing the ground for another possible research path 
which deserves a dedicated paper in the future.

\section*{Acknowledgements}
G.E. is grateful to INDAM for membership, and thanks the ET and QGSKY  
research units of INFN Naples.
M.R. acknowledges support from the INFN
Iniziativa Specifica QUAGRAP and from the European COST actions BridgeQG CA23130
and CaLISTA CA21109.
M.R. dedicates this paper to his sister Laura, and G.E. to his daughter Margherita.


\begin{thebibliography}{99}

\bibitem{HE1973}
S.W. Hawking and G.F.R. Ellis, {\it The Large Scale Structure of Space-Time}
(Cambridge University Press, Cambridge, 1973).
\bibitem{Chan1984}
S. Chandrasekhar, {\it The Mathematical Theory of Black Holes}
(Oxford University Press, Oxford, 1984).
\bibitem{WALD1974}
R.M. Wald, Gedanken experiments to destroy a black hole, 
{\it Ann. Phys. (N.Y.)} {\bf 82}, 548-556 (1974).
\bibitem{JW1977}
P.S. Jang and R.M. Wald, The positive energy conjecture and the cosmic 
censor hypothesis, {\it J. Math. Phys.} {\bf 18}, 41-44 (1977).
\bibitem{HPS2016}
S.W. Hawking, M.J. Perry and A. Strominger, Soft hair on black holes,
{\it Phys. Rev. Lett.} {\bf 116}, 231301 (2016).
\bibitem{Persides1976a}
S. Persides, Global properties of radial wave functions in Schwarzschild's
space-time. I: The regular singular point, {\it Commun. Math. Phys.}
{\bf 48}, 165-189 (1976).
\bibitem{Persides1976b}
S. Persides, Global properties of radial wave functions in Schwarzschild's
space-time. II: The irregular singular point, {\it Commun. Math. Phys.}
{\bf 50}, 229-239 (1976).
\bibitem{Frolov2017kze}
V.P. Frolov, P. Krtous and D. Kubiznak, Black holes, hidden symmetries,
and complete integrability, {\it Living Rev. Rel.} {\bf 20}, 6 (2017).
\bibitem{ER}
G. Esposito and M. Refuto, New perspectives on the irregular singular 
point of the wave equation for a massive scalar field in Schwarzschild
space-time, {\it Symmetry} {\bf 17}, 922 (2025).
\bibitem{Hamaide2023}
L. Hamaide and T. Torres, Black hole information recovery from gravitational
waves, {\it Class. Quantum Grav.} {\bf 40}, 085018 (2023).
\bibitem{Ronveaux1995}
A. Ronveaux and F.M. Arscott (eds.), {\it Heun's Differential Equations}
(Clarendon Press, Oxford, 1995).
\bibitem{Poincare1}
H. Poincar\'e, Sur les \'equations lin\'eaires aux diff\'erentielles ordinaires
et aux diff\'erences finies, {\it Am. J. Math.} {\bf 7}, 203-258 (1885).
\bibitem{Poincare2}
H. Poincar\'e, Sur les int\'egrales irreguli\`eres des \'equations lin\'eaires,
{\it Acta Math.} {\bf 8}, 295-344 (1886).
\bibitem{Dieu1980}
J. Dieudonn\'e, {\it Calcul Infinit\'esimal} (Hermann, Paris, 1980).
\bibitem{DEV2014}
E. Di Grezia, G. Esposito, and P. Vitale, Self-dual road to noncommutative
gravity with twist: A new analysis, {\it Phys. Rev. D} {\bf 89}, 064039 (2014). 
\bibitem{Wilc}
E.J. Wilczynski, {\it Projective Differential Geometry of Curves and 
Ruled Surfaces} (Teubner, Leipzig, 1906).
\bibitem{Esposito2017}
G. Esposito, {\it From Ordinary to Partial Differential Equations},
Unitext Series {\bf 106} (Springer Nature, Cham, 2017).
\bibitem{SCH1916}
K. Schwarzschild, \"{U}ber das gravitationsfeld eines massenpunktes nach
dei einsteinschen theorie, {\it Sitz. K. Preuss. Akad. Wiss.} 
189-196 (1916).
\bibitem{REISSNER1916}
H. Reissner, \"{U}ber dir eigengravitation des elektrischen feldes nach
der einsteinschen theorie, {\it Ann. der Physik} {\bf 355}, 106-120 (1916).
\bibitem{NORDSTROM1918}
G. Nordstr\"{o}m, On the energy of the gravitation field in Einstein's theory,
{\it K. Ned. Akad. Wet. Proc. Ser. B Phys. Sci.} 
{\bf 20}, 1238-1245 (1918).
\bibitem{Kerr1963}
R.P. Kerr, Gravitational field of a spinning mass as an example of 
algebraically special metrics, {\it Phys. Rev. Lett.} 
{\bf 11}, 237 (1963).
\bibitem{Newman1965}
E.T. Newman, E. Couch, K. Chinnapared, A. Exton, A. Prakash, and
R. Torrence, Metric of a rotating, charged mass, {\it J. Math. Phys.}
{\bf 6}, 918-919 (1965).
\bibitem{Mazur1982}
P.O. Mazur, Proof of uniqueness of the Kerr-Newman black hole solution,
{\it J. Phys. A Math. Gen.} {\bf 15}, 3173 (1982).
\bibitem{Carter1968}
B. Carter, Global structure of the Kerr family of gravitational fields,
{\it Phys. Rev.} {\bf 174}, 1559 (1968).
\bibitem{Wald1972}
R.M. Wald, Electromagnetic fields and massive bodies, 
{\it Phys. Rev. D} {\bf 6}, 1476 (1972).
\bibitem{NJ1965}
E.T. Newman and A.I. Janis, Note on the Kerr spinning-particle metric,
{\it J. Math. Phys.} {\bf 6}, 915-917 (1965).
\bibitem{KS1965}
R.P. Kerr and A. Schild, Some algebraically degenerate solutions of 
Einstein's gravitational field equations, in {\it Proceedings of
Symposia in Applied Mathematics} {\bf 17}, 199-209, 
Am. Math. Soc. (1965).
\bibitem{DKS1969}
G.C. Debney, R.P. Kerr, and A. Schild, Solutions of the Einstein and
Einstein-Maxwell equations, {\it J. Math. Phys.} 
{\bf 10}, 1842-1854 (1969).
\bibitem{AN2014}
T. Adamo and E.T. Newman, The Kerr-Newman metric: a review, 
arXiv:1410.6626 (2014).
\bibitem{BL1967}
R.H. Boyer and R.W. Lindquist, Maximal analytic extension of the
Kerr metric, {\it J. Math. Phys.} {\bf 8}, 265-281 (1967).
\bibitem{Wald1984}
R.M. Wald, {\it General Relativity} (University of Chicago Press,
Chicago, 1984).
\bibitem{KS2009}
R.P. Kerr and A. Schild, Republication of: A new class of vacuum
solutions of the Einstein field equations, 
{\it Gen. Relativ. Gravit.} {\bf 41}, 2485-2499 (2009).
\bibitem{LYNDEN2004}
D. Lynden-Bell, Electromagnetic magic: The relativistically 
rotating disk, {\it Phys. Rev. D} {\bf 70}, 105017 (2004).
\bibitem{KAISER2004}
G. Kaiser, Distributional sources for Newman's holomorphic Coulomb
field, {\it J. Phys. A Math. Gen.} {\bf 37}, 8735 (2004).
\bibitem{Teukolsky2015}
S.A. Teukolsky, The Kerr metric, {\it Class. Quantum Grav.}
{\bf 32}, 124006 (2015).
\bibitem{BCHR2014}
C.L. Benone, L.C. Crispino, C. Herdeiro, and E. Radu,
Kerr-Newman scalar clouds, {\it Phys. Rev. D} 
{\bf 90}, 104024 (2014).
\bibitem{FN2004}
H. Furuhashi and Y. Nambu, Instability of massive scalar fields in
Kerr-Newman spacetime, {\it Prog. Theor. Phys.}
{\bf 112}, 983-995 (2004).
\bibitem{BE1953}
H. Bateman and A. Erd\'elyi, {\it Higher Transcendental Functions. II}
(McGraw-Hill, New York, 1953).
\bibitem{NIST}
{\it NIST Digital Library of Mathematical Functions}.
https://dlmf.nist.gov/, Release 1.2.4 of 2025-03-15. Edited by
F.W.J. Olver, A.B. Olde Daalhuis, D.W. Lozier, B.I. Schneider,
R.F. Boisvert, C.W. Clark, B.R. Miller, B.V. Saunders, H.S. Cohl,
and M.A. McClain.
\bibitem{VBM2014}
H.S. Vieira, V.B. Bezerra, and C.R. Muniz, Exact solutions of the
Klein-Gordon equation in the Kerr-Newman background and Hawking
radiation, {\it Ann. Phys. (N.Y.)} {\bf 350}, 14-28 (2014).
\bibitem{VBM2022}
H.S. Vieira, V.B. Bezerra, and C.R. Muniz, Instability of the
charged massive scalar field on the Kerr-Newman black hole
spacetime, {\it Eur. Phys. J. C} {\bf 82}, 932 (2022).
\bibitem{RW1977}
D.J. Rowan and G. Stephenson, The Klein-Gordon equation in a
Kerr-Newman background space, {\it J. Phys. A Math. Gen.}
{\bf 10}, 15 (1977).
\bibitem{Chandra1976}
S. Chandrasekhar, The solution of Dirac's equation in Kerr
geometry, {\it Proc. R. Soc. Lond. A} {\bf 349}, 571-575 (1976).
\bibitem{Page1976}
D.N. Page, Dirac equation around a charged, rotating black hole,
{\it Phys. Rev. D} {\bf 14}, 1509-1510 (1976).
\bibitem{Kraniotis2019}
G.V. Kraniotis, The massive Dirac equation in the Kerr-Newman-de 
Sitter and Kerr-Newman black hole spacetimes, 
{\it J. Phys. Comm.} {\bf 3}, 035026 (2019).
\bibitem{Dariescu2021}
C. Dariescu, M.A. Dariescu, and C. Stelea, Dirac equation on the
Kerr-Newman spacetime and Heun functions, 
{\it Adv. H. En. Phys.} {\bf 2021}, 5512735 (2021).
\bibitem{DeWitt1984}
B.S. DeWitt, The spacetime approach to quantum field theory,
in {\it Relativity, Groups and Topology II: Les Houches Summer School on Theoretical Physics},
eds. B.S. DeWitt and R. Stora, pp. 381-738
(North-Holland, Amsterdam, 1984).
\bibitem{DeWitt2003}
B.S. DeWitt, {\it The Global Approach to Quantum Field Theory}, Vols. 1 and 2,
Oxford Series of Monographs on Physics, Vol. 114 
(Clarendon Press, Oxford, 2003).
\bibitem{Teukolsky1973I}
S.A. Teukolsky, Rotating black holes: Separable wave equations for
gravitational and electromagnetic perturbations, 
{\it Phys. Rev. Lett.} {\bf 29}, 1114-1118 (1972).
\bibitem{Teukolsky1973II}
S.A. Teukolsky, Perturbations of a rotating black hole. I. 
Fundamental equations for gravitational, electromagnetic, and
neutrino-field perturbations, {\it Ap. J.} {\bf 185},
635-648 (1973).
\bibitem{PT1973}
W.H. Press and S.A. Teukolsky, Perturbations of a rotating
black hole. II. Dynamical stability of the Kerr metric,
{\it Ap. J.} {\bf 185}, 649-674 (1973).
\bibitem{Chen:2022rrg}
C.-Y. Chen, X.-H. Wang, Y. You, D.-S. Sun, F.-L. Lu and S.-H. Dong,
Exact solutions to the angular Teukolsky equation with $s \neq 0$,
{\it Commun. Theor. Phys.} {\bf 74}, 115001 (2022).
\bibitem{Compere2018}
G. Comp\`ere and A. Fiorucci,
Advanced Lectures on General Relativity,
arXiv:1801.07064 [hep-th] (2018).
\bibitem{Berti2009}
E. Berti, V. Cardoso and A. O. Starinets,
Quasinormal modes of black holes and black branes,
{\it Class. Quant. Grav.} {\bf 26}, 163001 (2009),
doi:10.1088/0264-9381/26/16/163001.
\bibitem{Berti2006}
E. Berti and V. Cardoso,
Quasinormal ringing of Kerr black holes. I. The Excitation factors,
{\it Phys. Rev. D} {\bf 74}, 104020 (2006),
doi:10.1103/PhysRevD.74.104020.
\bibitem{Decanini2009}
Y. Decanini and A. Folacci,
Regge poles of the Schwarzschild black hole: A WKB approach,
{\it Phys. Rev. D} {\bf 81}, 024031 (2010),
doi:10.1103/PhysRevD.81.024031.
\bibitem{Leaver1985}
E.W. Leaver, An analytic representation for the quasi-normal modes of Kerr black holes,
{\it Proc. Roy. Soc. Lond. A} {\bf 402}, 285--298 (1985).
\bibitem{Sucu:2026gro}
E. Sucu, Dirac perturbations and thermodynamic corrections in a $F(R)$-corrected black hole,
{\it Phys. Scripta} {\bf 101}, 155001 (2026).
\bibitem{Sucu:2026rsr}
E. Sucu, Dirac quasinormal modes of charged black holes in $f(R,T)$ gravity with nonlinear electrodynamics,
{\it Phys. Lett. A} {\bf 585}, 131627 (2026).
\bibitem{Sucu:2026hfc}
E. Sucu, Dirac quasinormal modes, quality factor and gravitational lensing in nonlinear 
electrodynamics black holes with Barrow entropy,
{\it Nucl. Phys. B} {\bf 1026}, 117421 (2026).
\bibitem{Sucu:2026andp}
E. Sucu, İ. Sakallı and Y. Sucu, Unified analysis of fermionic tunneling, Barrow-exponential 
entropy corrections, and plasma lensing in Carrollian Reissner-Nordström spacetime,
{\it Ann. Phys. (Berlin)} {\bf 538}, e00588 (2026).
\bibitem{Gecim2020}
G. Gecim, Quantum gravity correction to the thermodynamic quantities of the charged dRGT black hole,
{\it Turk. J. Phys.} {\bf 44}, 564--578 (2020).
\bibitem{Gecim:2013mfa}
G. Gecim and Y. Sucu, Tunnelling of relativistic particles from new type black hole in new massive gravity,
{\it JCAP} {\bf 02}, 023 (2013).
\bibitem{Huh}
H. Huh, Asymptotic properties of the massive scalar field in the
external Schwarzschild space-time. \emph{J. Geom. Phys.} {\bf 2008}, \emph{58}, 55.
\end{thebibliography}
\end{document}